\documentclass[manuscript]{acmart}
\usepackage{tabularx}
\newcolumntype{L}{>{\raggedright\arraybackslash}X}
\AtBeginDocument{%
  \providecommand\BibTeX{{%
    \normalfont B\kern-0.5em{\scshape i\kern-0.25em b}\kern-0.8em\TeX}}}

\setcopyright{rightsretained}
\copyrightyear{2023}
\acmYear{2023}
\acmDOI{XXXXXXX.XXXXXXX}

\acmPrice{15.00}
\acmISBN{978-1-4503-XXXX-X/18/06}

\begin{document}

%%
%% The "title" command has an optional parameter,
%% allowing the author to define a "short title" to be used in page headers.
\title{Practitioner Trajectories of Engagement with Ethics-Focused Method Creation}

%%
%% The "author" command and its associated commands are used to define
%% the authors and their affiliations.
%% Of note is the shared affiliation of the first two authors, and the
%% "authornote" and "authornotemark" commands
%% used to denote shared contribution to the research.
\author{Colin M. Gray}
\email{gray42@purdue.edu}
\author{Ikechukwu Obi}
\email{obii@purdue.edu}
\affiliation{%
  \institution{Purdue University}
  \streetaddress{401 N Grant Street}
  \city{West Lafayette}
  \state{Indiana}
  \country{USA}
  \postcode{47907}
}

\author{Shruthi Sai Chivukula}
\email{schivuku@iu.edu}
\affiliation{%
  \institution{Indiana University}
  \streetaddress{901 E 10th St Informatics West}
  \city{Bloomington}
  \state{Indiana}
  \country{USA}}
  
\author{Ziqing Li}
\email{li3242@purdue.edu}
\author{Thomas Carlock}
\email{tcarloc@purdue.edu}
\author{Matthew Will}
\email{will10@purdue.edu}
\author{Anne C. Pivonka}
\email{apivonka@purdue.edu}
\author{Janna Johns}
\email{johns20@purdue.edu}
\author{Brookley Rigsbee}
\email{mccull24@purdue.edu}
\affiliation{%
  \institution{Purdue University}
  \streetaddress{401 N Grant Street}
  \city{West Lafayette}
  \state{Indiana}
  \postcode{47907}
  \country{USA}
}

\author{Ambika R. Menon}
\email{321ambika@gmail.com}
\author{Aayushi Bharadwaj}
\email{aayushi.bharadwaj@gmail.com}
\affiliation{%
  \institution{Srishti Institute of Art, Design, and Technology}
  \city{Bangalore}
  \country{India}
}

%%
%% By default, the full list of authors will be used in the page
%% headers. Often, this list is too long, and will overlap
%% other information printed in the page headers. This command allows
%% the author to define a more concise list
%% of authors' names for this purpose.
\renewcommand{\shortauthors}{Gray, et al.}

%%
%% The abstract is a short summary of the work to be presented in the
%% article.
\begin{abstract}
  Design and technology practitioners are increasingly aware of the ethical impact of their work practices, desiring tools to support their ethical awareness across a range of contexts. In this paper, we report on findings from a series of co-design workshops with technology and design practitioners that supported their creation of a bespoke ethics-focused action plan. Using a qualitative content analysis and thematic analysis approach, we identified a range of roles and process moves that practitioners employed and illustrate the interplay of these elements of practitioners' instrumental judgment through a series of three cases, which includes evolution of the action plan itself, the ethical dilemmas or areas of support the action plan was intended to support, and how the action plan represents resonance for the practitioner that created it. We conclude with implications for supporting ethics in socio-technical practice and opportunities for the further development of ethics-focused methods that are resonant with the realities of practice.
\end{abstract}

%%
%% The code below is generated by the tool at http://dl.acm.org/ccs.cfm.
%% Please copy and paste the code instead of the example below.
%%


%%
%% Keywords. The author(s) should pick words that accurately describe
%% the work being presented. Separate the keywords with commas.

%%
%% This command processes the author and affiliation and title
%% information and builds the first part of the formatted document.
\maketitle

{\color{red}\textbf{Draft: September 13, 2022}}

\section{Introduction}

Scholars and practitioners alike are increasingly interested in creating, understanding, and supporting ethically-focused design and technology practices. At the center of this interest are numerous competing interests and epistemological stances, forms of complexity, and disciplinary approaches that frame ethics in different ways. Design and technology practitioners, their teams, and organizations they represent must consider what is ethical, for whom, and how they know---seeking to harmonize rapidly changing legal and regulatory standards~\cite{Gray2021-zf}, growing public concern over manipulative design practices~\cite{Bongard-Blanchy2021-wj}, and a dearth of definitive or broadly applicable standards in many technology professions that address pressing ethical issues ~\cite{Wong2020-ki,Buwert2018-uw,Debs2022-mt,Dindler2022-ny}. 

Over the past decade, numerous toolkits, resources, and methods have been proposed to support ethically-focused design practices, including academic methodology-driven efforts such as \textit{Value Sensitive Design} (VSD; ~\cite{Friedman2019-zg}) and \textit{Values at Play}~\cite{Flanagan2014-hf}; academic method-driven efforts such \textit{Judgment Call the Game}, \textit{GenderMag}~\cite{Burnett2016-dr}, and \textit{Speculative Enactments}~\cite{Elsden2017-lu}; and practitioner method/toolkit-driven efforts such as Kat Zhou's \textit{Design Ethically} toolkit or Jet Gispen's \textit{Ethics for Designers} toolkit~\cite{Chivukula2021-xk}. Across this increasingly complex ethical landscape that has been described by practice-led scholarship~\cite{Wong2020-ki,Lindberg2020-wk,Gray2019-ep,Dindler2022-ny}, it is recognized that practitioners need resources to support both ethical awareness and their ability to act, once they build the requisite levels of awareness. However, tools to support ethical awareness and action are generally not well known by practitioners, do not comprehensively address matters of ethical concern, or are otherwise not resonant or responsive to the felt complexity of everyday design and technology work~\cite{Chivukula2020-bv,Lindberg2020-wk,Lindberg2021-hi,Watkins2020-zr,Gray2019-ep}. 

In this study, we build upon interest by both practitioners and scholars in facilitating the creation of tools that are both appreciated by design practitioners as \textit{resonant} with the demands of their everyday practice~\cite{Goodman2011-ak,Stolterman2008-ho,Stolterman2008-ty}. We leverage traditions of method design and other forms of ethical support undertaken by practitioners and researchers~\cite{Chivukula2021-xk,Friedman2017-rd,Shilton2018-ws}, but explicitly shift the framing of method design from \textit{designing for} practitioners in a user-centered design tradition to facilitating spaces for \textit{practitioners to design methods themselves}. We created a virtual co-design environment which was used to lead 26 design and technology professionals (13 practitioners and 13 students with professional experience) through the process of creating their own bespoke ethics-focused ``action plan.'' Across six collaborative 180 minute virtual workshops conducted on Zoom and Miro, groups of 3--6 participants iteratively: 1) identified ethical dilemmas they faced in their work environment, 2) ``shopped'' for building blocks of existing ethics-focused methods they felt were relevant to addressing their dilemma, 3) built a prototype of an action plan to support their work using the building blocks they selected alongside other resources, and 4) evaluated their action plan, considering how it might be adapted for alternate use contexts. Through this process, each participant built a more detailed understanding of the ethical complexity of their practice, created a bespoke action plan to address that complexity, and in most cases, recognized an ability to make changes in their workplace in ways they had not fully appreciated prior to the workshop. We analyzed the outcomes of these workshops using a qualitative content analysis~\cite{Hsieh2005-ld} and reflexive thematic analysis approach~\cite{Braun2021-dt}, first identifying how participants took on a range of ``roles'' or attitudes towards their context that framed their ethical stance. Building on this analysis, we then identified how participants made their felt ethical complexity tractable through problem framing, using their component instrumental and appreciative judgments to characterize \textit{process moves} that they utilized to inform the final shape of their action plan. We then constructed detailed case studies of three diverse participants to further characterize the action plan design process and participants' engagement with their felt ethical complexity in achieving final outcomes. Through these forms of analysis and the resulting findings, we answer the following research questions:

\begin{enumerate}
    \item What roles and process moves do participants use to structure and navigate their ethics-focused method design process?
    \item How do participants engage the ethical complexity of their role as they created a bespoke ethics-focused action plan?
\end{enumerate}

The contribution of this paper is two-fold. First, we describe a range of ethics-focused roles that inform both method selection and design, indicating opportunities for further nuance in describing implementation of new ethics-focused approaches in practice and means of supporting the activation of ethics-focused knowledge. Second, we provide a thick description of practitioners' engagement in the method design process, providing insights into important aspects of resonance and ecological complexity that could be used to support future method design and practice-led research.

\section{Related Work}
To situate our contribution, we first describe why ethically-focused design has been so difficult to accomplish and the forms of design complexity that resist ``simple'' solutions. We then build on the notion of ethical design complexity, identifying instances where designers of ethics-focused methods have sought to support ethically-focused design practices, outlining a potential expansion of design knowledge by structuring practitioner-led efforts to support their own work contexts.

\subsection{Ethical Design Complexity and the Challenges of Supporting Ethically-Focused Practice}

There is a large and growing body of research that describes how design and technology practitioners engage with ethics and values as part of their everyday work experiences ~\cite{Gray2019-ep,Lindberg2020-wk,Chivukula2021-oj,chivukula2021identity,Lindberg2021-hi,Wong2021-pv,Boyd2021-sv,Dindler2022-ny, Lindberg2020-wk,Shilton2021-gy,Friedman2019-zg,Frauenberger2017-mk,Wong2020-ki}. Scholars have examined issues relating to ethics and ethically-focused design practices from numerous perspectives, including: characterizing the strategies practitioners employ to navigate ethical complexities within their organization ~\cite{Gray2019-ep,Lindberg2020-wk,Chivukula2021-oj,chivukula2021identity,Lindberg2021-hi,Wong2021-pv,Boyd2021-sv,Dindler2022-ny}; empowering practitioners with design methods and toolkits that resonate with their practice and support ethically-aware decision making~\cite{Lindberg2020-wk,Shilton2021-gy,Friedman2019-zg,Frauenberger2017-mk,Wong2020-ki}; introducing or expanding ethics education into the HCI curriculum as an approach to equip students and practitioners to handle ethical complexity~\cite{Slager2021-vv,G_Pillai2021-hm,Gray2021-gl,Loke2020-oe,Fore2019-vi}; and building accounts of how methods or tools can be developed to support practitioners~\cite{Gray2022-kv,Chivukula2021-xk,Gray2022-na,Gray2016-pa,Pivonka2022-nm}.

The concept of \textit{ethical design complexity} captures some of the key elements that make the work of practitioners in relation to ethics so difficult to manage, describe, and support, defined by Gray and Chivukula as ``the complex and choreographed arrangements of ethical considerations that are continuously mediated by the designer through the lens of their organization, individual practices, and ethical frameworks''~\cite{Gray2019-ep}. This articulation of complexity as ecologically situated builds upon a range of ethics scholarship which describes how practitioners engage in ethical decision-making and sense-making~\cite{Shilton2017-zu,Boyd2021-sv,Chivukula2020-bv,Lindberg2021-hi} and seek to make changes based on their profession or organizational role~\cite{Wong2021-pv,Shilton2013-dq,Lindberg2020-wk,Chivukula2021-oj,Watkins2020-zr}. 

HCI scholars have explored numerous ways to empower practitioners in navigating ethical complexity in their everyday practice. Lindberg et al.~\cite{Lindberg2020-wk} engaged with practitioners to explore ways of supporting them to integrate ethical values into their everyday practice. Findings from their research suggest that co-creation activities might be one of the best ways of helping designers to develop methods that resonate with the ethical complexities they encounter in their everyday practice. Shilton et al.~\cite{Shilton2021-gy} acknowledged that no single design method will be sufficient for resolving ethical complexities but that an amalgamation of ethical tools and methods will help to drive change from different facets towards ensuring an ethical culture. Wong et al.~\cite{Wong2020-ki} encouraged practitioners to go beyond the ``ethics checklist'' to explore how to use games, roleplaying, and critical making as means of integrating ethics into design practice. Frauenberger et al.~\cite{Frauenberger2017-mk} proposed the use of in anticipatory ethics to resolve ethical complexity in technology practice. More conceptually, Lindberg et al.~\cite{Lindberg2021-hi} also investigated practitioners' understanding of ethics. Findings from their research revealed that noticing, reflecting, and reacting were three dominant ways in which practitioners approach ethical issues within their organization. Tulloch et al.~\cite{Tulloch2019-dx} argued that design researchers must recognize their position within their organization's ethical ecology to be able to determine approaches that will support them to induce meaningful change. d'Aquin et al.~\cite{DAquin2018-bl} advocated for the need to include data scientists in the discourse around ethics and developed an ``Ethics by Design'' research methodology for conducting research in the fields of AI and Data Science. And Reijers \& Gordin~\cite{Reijers2019-sd} advocated for practitioners to transition from Value Sensitive Design (VSD) to Virtue Practice Design (VPD), arguing that while VSD focuses on the artifacts, VPD focuses on the process and agents enacting the design to ensure that they are virtuous. They remarked that the education of the practitioners plays a crucial role in fostering an ethical and virtuous design practice. Altogether, this range of scholarship illustrates different strands of practical and conceptual support within the HCI community to describe and seek to support practitioners in navigating ethical complexity in their everyday practice.

Although HCI scholars have studied different forms of ethical complexity and designed methods that practitioners may employ to navigate these challenges, little research has described how practitioners select, appropriate, adopt, and build methods that resonate with the particular ethical complexity relevant to their practice. As one rare example, Wong \cite{Wong2021-pv} investigated the strategies that user experience professionals employ to navigate the ethical complexity within their organization with the goal of inducing ethical outcomes. Findings from their study revealed that practitioners deploy those tactics to achieve three goals, including (1) advocating for the use of UX expertise in resolving those kinds of issues; (2) making their values visible within their organization; (3) altering their organizational processes to make it more ethical. Another example from Shilton~\cite{Shilton2013-dq} illustrates how not only formal methods can be used to encourage an ethical focus, but also a consideration of how organizational forces can be reshaped by creating ``values levers'' to take advantage of specific moments of awareness in ways that can shift organizational culture and the ability to act. These findings align with prior work from Gray, Chivukula, and colleagues that include descriptions of the tensions that UX practitioners face when seeking to address ethical issues in their workplace \cite{Watkins2020-zr}, the interplay of identity claims and forms of action that are individually mediated \cite{Chivukula2021-oj}, and dimensions of practice that can support ethically-focused action~\cite{Chivukula2020-bv}. In this paper, we seek to contribute to this growing body of research on ethical complexity in HCI focusing on characterizing the navigational maneuvers and roles practitioners employ as they create ethics-focused support tools that resonate with their own experience of design and technology practice. %unravelling this of this phenomenon by providing insights into ways ... which has received limited insight within the HCI community.

\subsection{(Ethics-Focused) Methods and Design Knowledge}
Numerous methods, toolkits, and other resources have been proposed to enable technology and design practitioners to address, evaluate, or develop alignment around ethical issues that impact their everyday work~\cite{Chivukula2021-xk,Friedman2017-rd,Shilton2018-ws}. Methodologies driven by moral philosophy such as \textit{Value Sensitive Design}~\cite{Friedman2019-zg} are likely the best known in scholarly and educational contexts, while practitioners often rely upon toolkits or resources that are oriented more towards specific contexts of use (e.g., the EthicalOS Toolkit~\footnote{\url{https://ethicalos.org}}), technologies (e.g., Microsoft's \textit{Guidelines for AI Interaction}~\cite{Amershi2019-oh}), or values (e.g., Microsoft's Inclusive Design Toolkit~\footnote{\url{https://www.microsoft.com/design/inclusive/}}). As scholars have previously found, monolithic toolkits or methods are often not resonant with the realities of everyday practice~\cite{Gray2016-pa,Goodman2011-ak} and the ethical design complexity felt by practitioners involves the mediation of many forces which which cannot always be considered in advance. Thus, our focus in this paper was to scaffold practitioners' ability to create their own support tools, using their knowledge of their work environment along with ``building blocks'' of existing tools to support ethical awareness and action in ways that were salient to them. 

To frame these support tools and scaffolding through co-design, we leverage existing concepts from the design theory literature and prior work that describes how method designers create new methods that allow us to analyze practitioners' design processes as they create their bespoke ethics-focused toolkit. Designers continuously make complex and layered judgments that inform their understanding and operationalization of the problem space and facilitate their engagement in design work~\cite{Gray2015-qi,Nelson2012-ov,Schon1984-oe}, and in the context of method design, creative constraints~\cite{Biskjaer2014-gy} are actively used to shape the problem space and consider potential impact~\cite{Gray2022-kv}. Nelson and Stolterman~\cite{Nelson2012-ov} describe a set of eleven judgment types which have been operationalized in further empirical work, and in this paper we focus on a subset of judgment types which are particularly impactful in the design of a ethics-focused toolkit, including: instrumental, appreciative, and framing judgments. \textit{Instrumental judgments} refer to ``the capacity to choose appropriate approaches to design problems, decide from an array of 
established options, or create new approaches''~\cite{Murdoch-Kitt2020-sw} with a focus on which tools and methods the designer selects, and through what capabilities these tools or methods are operationalized. \textit{Appreciative judgments} refer to the ``[p]lacing [of] high value and emphases on certain aspects of a design situation while backgrounding, or lessening focus on  others''~\cite{Gray2015-qi}, whereby designers use an appreciative system---or ``normative framing of the situation''~\cite{Schon1984-oe}---to make sense of the design situation in ways that value certain kinds of facets and end states. Finally, \textit{framing judgments} refer to the introduction of constraints to make the problem space tractable ``starting from the only `known' in the equation, the desired value, and then adopting or developing a frame that is new to the problem situation,''~\cite{Dorst2015-aq} thereby ``[c]reating a working area for design activities to occur''~\cite{Gray2015-qi}. Across these three judgment types, we would expect issues relating to ecological resonance and the value orientations of the designer and organization to primarily be addressed through \textit{appreciative} judgments; \textit{framing} judgments support the identification of a tractable design space with constraints relating to number(s) and type(s) of actors and specific goals that the designer wishes to support; and \textit{instrumental} judgments articulate to what degree a tool is likely to be relevant or useful in the everyday work practices of the designer while advancing the goals articulated through appreciative and framing judgments. 

\begin{figure*}[!htb]
    \centering
    \includegraphics[width=\textwidth]{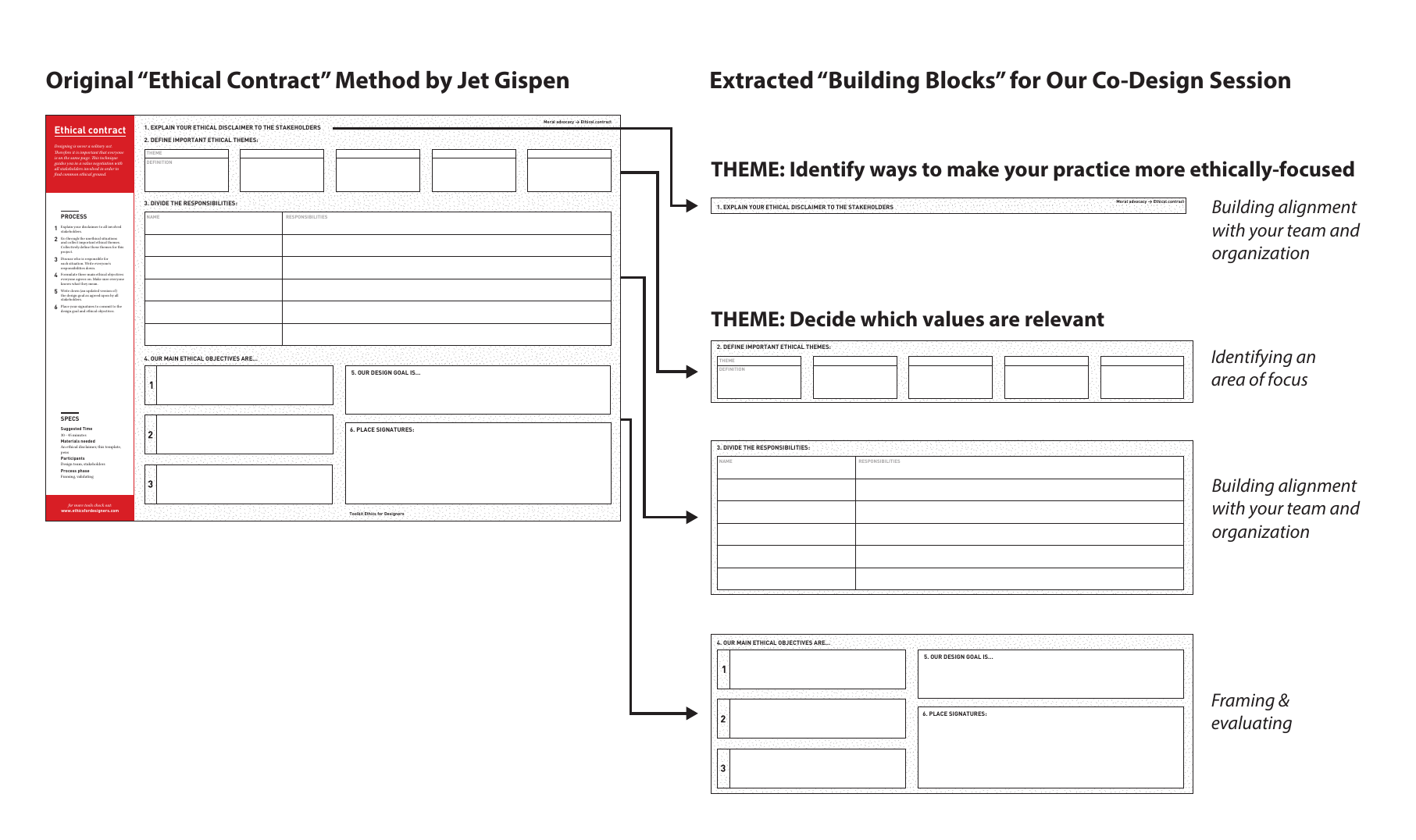}
    \caption{Example of ethics-focused method decomposition into building blocks, including organization of the resulting blocks into \textbf{themes} (bolded) and \textit{use cases} (italicized). The \textit{Ethical Contract} method is originally created by Jet Gispen~\cite{ethicalcontract}}
    \label{fig:methoddecomposition}
\end{figure*}

In our co-design study, we asked technology and design practitioners and students to engage with existing design knowledge in the form of method ``building blocks'' that would then inform their creation of a bespoke ethics-focused toolkit. The idea behind these ``building blocks'' was inspired by Woolrych et al.'s~\cite{Woolrych2011-db} observation that methods are not used as ``indivisible wholes,'' but rather can be considered as ``ingredients'' that can be used by designers to form many different ``meals.'' Complementary to this approach to methods is Gray's articulation of method ``cores''~\cite{Gray2022-na,Gray2016-pa}, which refer to ``the central conceit or framing metaphor that makes the entire method, or a portion of the method, coherent and potentially interchangeable.'' In order to maximize the flexibility of existing methods and strengthen the ``ingredient'' metaphor for our participants, we selected a subset of methods from a larger set of 63 ethics-focused methods from a collection proposed by Chivukula et al.~\cite{Chivukula2021-xk}, seeking to identify a wide range of building blocks. In total, we drew from 11 methods and created 73 total building blocks in building the items contained within the ``shop'' floor of the virtual co-design space (see Figure~\ref{fig:methoddecomposition} for an example of this method decomposition process).

\section{Our Approach}
In this co-design study, we engaged a range of technology and design practitioners in a series of interactive activities through a 180 minute virtual session. Through these activities, the practitioners iteratively identified an ethical dilemma they faced in their everyday work, selected relevant components of existing ethics-focused methods, and used these components to construct their own bespoke ethics-focused toolkit. Across these interactions, we mapped each practitioner's trajectory of engagement in relation to their area of desired ethical impact, including both iterative toolkit drafts and the process moves that shaped the intermediate and final toolkits.

%\subsection{Research Context}
%1. Our research context is situated and motivated by the intersection of the increasing manifestation of unethical design artifacts in various platforms; the concomitant increase in the number of ethics-focused methods designed to support practitioners navigating these complexities and challenging scenarios; and the oft report of lack of resonance between the developed methods and realities and dynamism of everyday design practice. 

%2. Achieving method-to-practice resonance of ethics-focused design methods and tool-kits is crucial to fostering an ethical and value-driven socio-technical practice 

%3. Considering the restrictions of physical contact during the pandemic, this study was held via virtual platforms that included Zoom and Miro. However, this restriction also presented us with an opportunity and allowed us to recruit participants that were outside of our immediate vicinity, and thus, enriching the pool of participants for the study in the process. 

%4. Our goal for this research is to shed light on some of the navigational maneuvers and roles/identities practitioners adopt to be able to effectively position themselves and employ methods and tool kits to resolve the ethical complexities they encounter in their everyday practice.

\subsection{Sampling Strategy}
We used a stratified sampling approach to build sets of participants for six co-creation sessions, with the strata including current role in design and technology work (student or practitioner), years of experience, industry type, and primary professional role (UX Designer, UX Researcher, Product Manager, Data Scientist, Data Engineer, and Software Engineer). To identify participants, we circulated a recruitment screener on a range of social media platforms, including Twitter, LinkedIn, and Reddit, as well as the professional networks of members of the research team. The inclusion criteria for participating in the co-creation sessions were structured separately for industry practitioners and students. For industry practitioners, the inclusion criteria included current employment in a design or technology-related role in industry with one or more years of experience. For students, the inclusion criteria included some form of past industry experience, such as a professional internship, and student participants primarily included those training to become UX designers and product managers. Since our co-creation objective was to empower the participants to identify and seek to address an ethical dilemma they have encountered in their professional practice, our criteria excluded any applicant that had no industry experience from participating in the sessions. In all, our sampling strategy produced a diverse group of practitioners and students from different professional roles. 

\subsection{Participants}

We conducted six co-creation sessions with 26 participants, including a total of 13 practitioners and 13 students. Three sessions were held with practitioners and three sessions were held with student participants . The first practitioner session had four participants, including two UX Designers, a Product Manager and a UX Researcher. The second practitioner session had three participants including two UX Designers and a UX Research Lead. The third practitioner session involved six practitioners, including a Product Manager, Data Scientist, Data Engineer, Software Engineer, and two UX Designers. Across the three practitioner sessions, participants had an average of 3.7 years of experience (SD=2.6; Min=1; Max=10) and worked for a range of company types, including Agencies or Consultancies, Enterprise (B2B), and Enterprise (B2B2C). The first student session included four participants: a UX Researcher and three UX Designers. Four participants also participated in the second student co-creation session, including a Product Manager and three UX Designers. The last student co-creation session involved five participants and included a Product Manager and four UX Designers. Before the sessions, all participants (both student and practitioners) were assigned with a unique identifier and icon to navigate the sessions pseudononymously if they chose to do so.

\subsection{Data Collection}
\begin{figure*}[!htb]
    \centering
    \includegraphics[width=\textwidth]{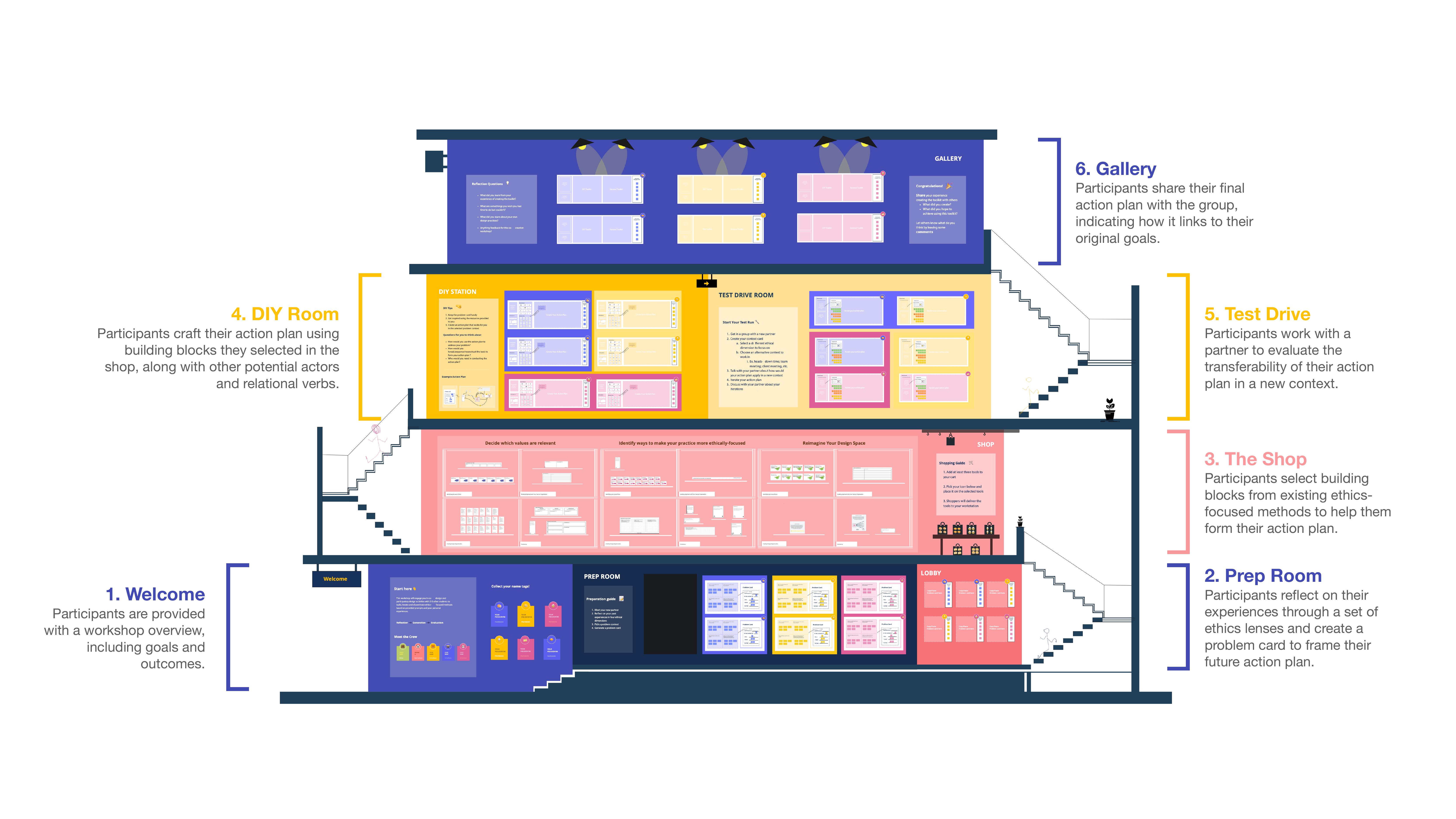}
    \caption{Co-creation session experience on Miro with six activities organized across four ``floors.''}
    \label{fig:cocreationhouse}
    \Description[Co-Creation Workshop Workspace]{An artifact from the co-creation workshop which is a representation of a house including four main floors with stairs leading up to each higher level, each level being a stage in the workshop. The bottom floor contains titles starting with "Start Here," "Collect your name tags!" "Prep Room" and "Lobby." Following that, a representation of a staircase leads to the second floor. The second floor contains titles starting with "Shop" and "Shopping Guide" and is followed by three more titles that say, "Reimagine Your Design Space," "Identify ways to make your practice more ethically-focused," and "Decide which values are relevant." The last three titles label three quadrants containing elements for the workshop. Following that, the representation of a staircase leads to the third floor. The third floor is divided into two halves, one side being titled, "DIY Station" and the other side being titled, "Test Drive Room." Six work spaces are contained on each half. Following that, the representation of a staircase leads to the fourth floor, which contains the six work spaces which sit under spot lights. This final floor is titled, "Gallery."}
\end{figure*}

The co-creation sessions were hosted on Zoom using breakout rooms and on Miro, a digital whiteboard platform. The final co-creation experience was visually organized as a virtual ``house'' on Miro (Figure~\ref{fig:cocreationhouse}) containing four ``floors,'' intended to foster an interactive and collaborative co-creation experience that stimulated the participants to collaborate, brainstorm, and work towards developing an action plan that would help them address the ethical complexities they experience in their everyday practice. We relied upon multiple facilitators, who used breakout rooms to support  different groupings of participants in interacting with each other across the session. The overall structure, along with questions or prompts participants were asked to consider and relevant collected data, is detailed in Table~\ref{table:data}. The first floor (Figure~\ref{fig:cocreationhouse}, \#1 and \#2) was designed to facilitate introductions and reflections on ethical dilemmas the practitioners intend to address. The second floor (Figure~\ref{fig:cocreationhouse},~\#3) was designed as a shopping area where the participants could shop for different ethics-focused methods that they could use to create an action plan that supported the participant in addressing the ethical dilemmas they identified on the first floor. We populated this floor with a set of 73 method ``building blocks,'' deconstructed from a set of existing ethics-focused methods and curated to provide a range of ``cores'' to support different kinds of toolkits and ethical dilemmas. Sample ``aisles'' of the shop included \textit{intention themes} (e.g., ``identify ways to make your practice more ethically-focused'', ``reimagine your design space'') and ``shelf areas'' within these aisles contained bundles of blocks organized by \textit{action orientations} (e.g., ``evaluating'', ``creating design opportunities'', ``building alignment with your team and organization''). The third floor (Figure~\ref{fig:cocreationhouse},~\#4 and \#5) was designed as a DIY workspace where participants used the methods they selected from The Shop to design an action plan. After the initial action plans were created, participants were paired with a new participant in a new breakout room to evaluate their method and identify how it would need to be altered to address a new context. The fourth floor (Figure~\ref{fig:cocreationhouse},~\#6) was designed as a gallery space where the participants could share and reflect on the action plan they created. Altogether, the co-creation sessions consisted of a series of activities designed to last cumulatively for three hours, including: 15 minutes of introductory and preparatory activities, 20 minutes for reflection and idea generation, 5 minutes for feedback on ideas, 10 minutes for shopping for methods to resolve identified problems, 30 minutes for developing an action plan to resolve identified challenges, 25 minutes for testing their action plan in a different context and iterating on their plan, and 10 minutes for reflecting on their experience during the session.

\begin{table*}[!htb]
    \begin{tabularx}{\textwidth}{LXr}
       \toprule
       \textbf{{Co-Creation Stage \newline [Space from Figure~\ref{fig:cocreationanalysisexample}]}} & \textbf{Questions} & \textbf{Data Collected} \\
       \midrule
    \textit{Introduction [Welcome Lobby]} & a) Can you tell us the name you would like to go by during the session? b) your industry role? and c) what you’re looking forward to in this workshop? & Audio\\
    \textit{Reflection \& idea generation} & a) What are some of the ethical dilemmas you have experienced? b) what are some of the situations or contexts in which you felt uncomfortable as a result of an ethical issue? c) what are some things that you wish you could do but are unable to for any reason? d) And lastly we ask that you consider any advancements in the field or future consequences that you may have concerns about & Text \& audio\\

    \textit{Problem space [Prep Room]} & Can you tell us about the problem card you created? & Audio \& card\\

    \textit{Developing an action plan [DIY Room]} & a) What are you thinking of making? b) Are there any difficulties you are facing in creating this action plan? c) Do you want any feedback from your partner? & Audio \& artifacts\\
    
    \textit{Testing the plan [Test Drive Room]} & Can you walk us through how would you go about applying this action plan in the selected problem context? & Audio \& refined artifacts\\
    
    \textit{Final Reflection[Gallery]} & a) What did you learn from your experience of creating your action plan? b) What are some things you wish you had time to do but couldn't? c) what are the things you learned about your own design practices? & Audio \\

   \bottomrule 
     \end{tabularx}
     \caption{Probing questions and data collected during each stage of the co-creation session.}
     \label{table:data}
 \end{table*}

\subsection{Data Analysis}
We commenced our data analysis by transcribing the video and audio produced during the co-creation session into text using Dovetail, a qualitative data analysis software tool. We then duplicated the artifacts created during the co-creation sessions on Miro into a new Miro board to allow for data analysis and comparison across sessions while preserving the original content. We then analyzed the data collaboratively through multiple stages, employing qualitative content analysis, role analysis, thematic analysis, and case study analysis. All analysis stages involved six researchers, including the principal investigator, a graduate student, and five undergraduate students. All researchers were trained in qualitative analysis and had prior experience working on qualitative, ethics-focused research projects. The data analysis steps for this research included familiarizing with the data, journey mapping, qualitative content analysis, role analysis, thematic analysis, and case study analysis. In the subsections below, we describe the activities conducted during each of the stages, including: familiarizing ourselves with the data, creating artifact-focused journey maps, and our use of thematic analysis to describe the roles and process moves of the participants.

\subsubsection{Familiarizing Ourselves with the Data through Content Analysis}

We began by familiarizing ourselves with the artifacts generated by the participants during the co-creation sessions. Through this process, we sought to sensitize ourselves with the content of the entire dataset in preparation for a robust data analysis---in some cases, reflecting on sessions we had facilitated and in other cases engaging with data collected with other facilitators for the first time. In this process of sensitization, we sought to identify the issues the participants came to the session hoping to address, the method building blocks they selected to design an action plan that responded to those issues, and the kinds of changes they made when iterating on their action plan. When engaging with these data, all researchers applied preliminary codes to the artifacts that related to our research questions using a qualitative content analysis approach~\cite{Hsieh2005-ld}. The researchers then discussed the codes generated from this exercise and reflected on their different interpretations of the data. Across data from all six co-creation sessions, we found that the participants sought to design an action plan to help them accomplish a range of different objectives, including: disseminating and fostering ethical awareness within their organization or team; changing a current process within their organization, while implicitly characterizing existing processes as unethical; or focusing on a small yet urgent ethical issues within the context of their practice that they believed need to be addressed. We also found that the participants employed multiple strategies to design their action plan, including a reframing or operationalization of their ethical concerns to make them tractable. Based on these initial findings, we decided to use a combination of reflexive thematic analysis and role analysis---using a visual journey map to ground the trajectories of participants in the sequence of co-creation activities that supported the design of their ethics-focused action plan.

\begin{figure*}[!htb]
    \centering
    \includegraphics[width=\textwidth]{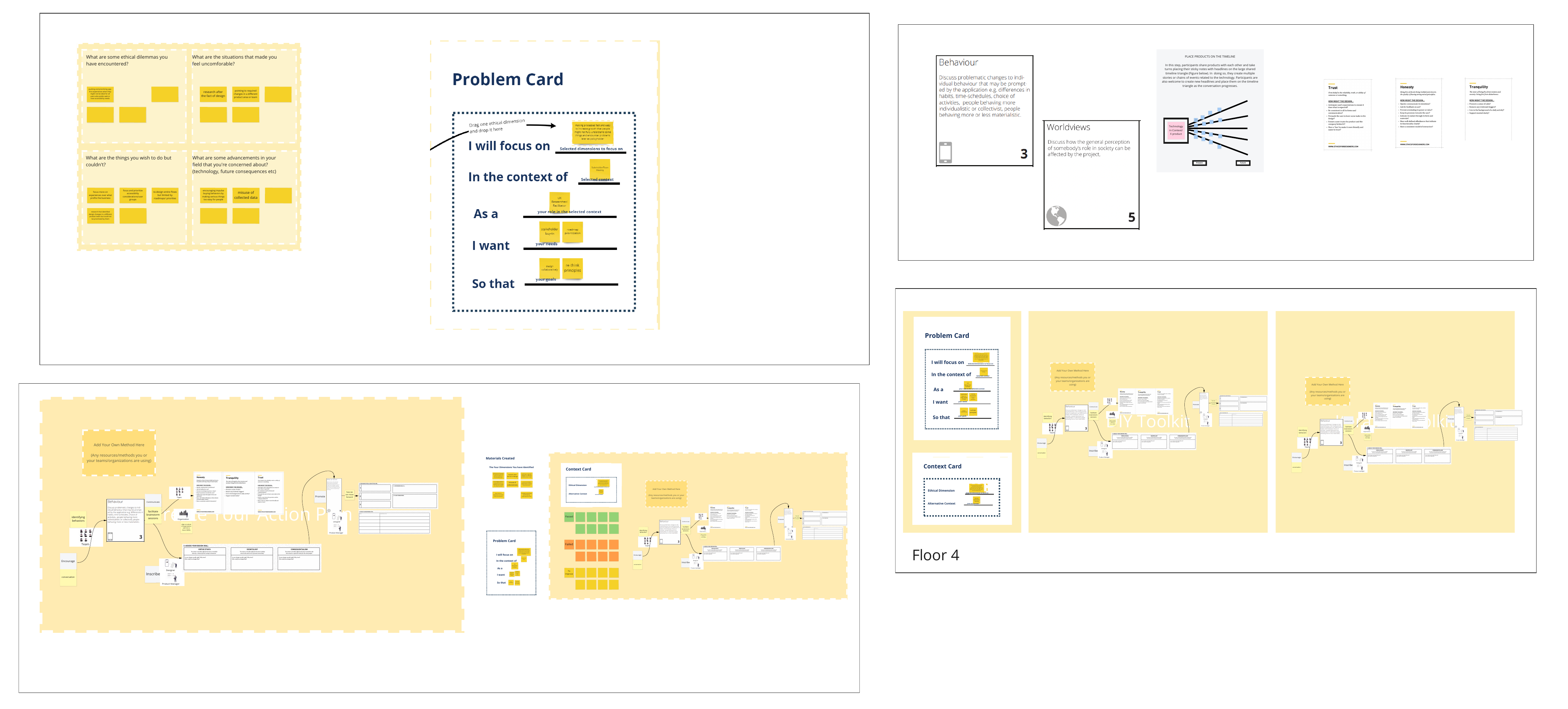}
   % \subfigure[]{\includegraphics[width=0.45\textwidth]{Figures/Interactional Qualities/IQ-Sociality.jpg}}
    \caption{An example of an artifact-focused journey map for a single co-design participant.}
    \label{fig:cocreationanalysisexample}
    \Description [Artifacts from one participant's co-creation experience.]{Four screenshots from practitioner's engagement with the Co-Creation workshop in Miro. First screenshot contains a workspace with prompts and sticky notes filled in by the participant and is next to an item titled, "Problem Card" containing prompts with blank spaces, the prompts being, "I will focus on," "In the context of," "As a," "I want," and "So that" and following each are the practitioner's answers on sticky notes. The next screenshot contains six artifacts. The first is titled, "Behavior" followed by the definition and a phone symbol, the second is titled and is titled, "Behaviors" followed by the definition and a world symbol, the third is titled, "Place Objects on the Timeline" and it also followed by a definition and blank space that has lines pointing away and blank spaces on the line, and the last three elements are titled, "Trust," "Honesty," and "Tranquility" and each contain a definition. The next screenshot is an overview of the practitioner's action plan in which includes the artifacts, more sticky notes, and other building blocks, all connected by arrows. A second workspace is visible which contains a new card titles, "Context Card." The last screen shot contains both the "Problem Card" and the "Context Card" and the same two filled in work spaces.}
\end{figure*}

\begin{table*}[!htb]
    \begin{tabularx}{\textwidth}{p{.15\textwidth}X}
       \toprule
       \textbf{Ethical Role} & \textbf{This role\ldots} \\
       \midrule
    \textit{The Advocate} & seeks to take action based on an intrinsic interest and awareness of their ethical role within their organization. Their focus in building their action plan through this role is to translate their personal awareness to others within their organization.\\

    \textit{The Reformer} & recognizes that their vision of ethically-focused action is not equally shared by others in their organization. Their focus in building their action plan through this role is to activate their intrinsic desire for ethical change in ways that might effect substantial change at the organizational or professional role level.\\

    \textit{The Operationalizer} & identifies a component of their practice and experience that relates to ethical awareness, but does not situate that knowledge in relation to broader ecological complexities that might have given rise to those ethical issues. Their focus in building their action plan involves honing a small ``piece of the puzzle’’ without addressing ecological implications of the proposed plan.\\

   \bottomrule 
     \end{tabularx}
     \caption{Ethics-focused roles that participants employed to navigate the creation of their action plan.}
     \label{table:roles}
 \end{table*}

\subsubsection{Identifying Participant Trajectories through Journey Mapping}
Building on our reflections from our preliminary analysis, we used an artifact-focused journey mapping approach to trace, characterize, and analyze the trajectory of each participant during the co-creation sessions. By collecting all artifacts created by each participant as one collection, we were able to trace the ethical problems the participants listed in their problem cards at the beginning of the session, the ethical challenge they elected to focus on during the session, the ethics-focused methods items they picked for designing their action-plan, the final action plan they created from those ethics-focused methods, and their reflection at the end of the session on their rationale for creating the action plan (see an example of this collection in Figure~\ref{fig:cocreationanalysisexample}). We also paired the artifacts created on the Miro Board with the video and text transcript produced from the sessions to describe the rationale provided by the participants for the different actions they took during the session, reflexively engaging with the following sensemaking questions:  1) How would I characterize the trajectory of this participant?; 2) What are the qualities within this trajectory that appear to be especially interesting or pivotal? 3) Does the participant engage in an iterative refinement of their concerns as compared to where they started? 4) Does the participant re-characterize or re-frame their initial ethical concern as they are confronted with these new tools? 5) Do participants find resonance between their ethical challenges and the ethics-focused method building blocks, or do they experience a misalignment between their objectives and the method building blocks? 6) How does the participant frame technical portions of their work as ethical (or not)? As we went through this analysis process, we individually produced memos to characterize the trajectory, roles, and process moves made by the participant both in setting up their problem space and throughout their participation in the sessions. After this individual analysis, we evaluated each participant journey map as pairs to reflectively engage with the interpretations of others in the research team---an important early recognition of our reflexive engagement as individual researchers and as a team.

\subsubsection{Characterizing Participant Trajectories through Role and Thematic Analysis}
Findings from our data sensitization and journey mapping analysis revealed that the participants took on multiple roles in the session to navigate the ethical complexity of their practice context which impacted the development of their action plans. We also found that the participants engaged in navigational maneuvers that allowed them to overcome perceived constraints or to take advantage of newly emerging opportunities for impact during the creation of their action plan. We continued our analysis of the materials generated by each participant during the co-creation session based on these points of departure, using role analysis~\cite{Carspecken1996-bq,Gray2021-gl,chivukula2021identity} to characterize the stance(s) participants took towards their action plan development process and its relationship to their felt ethical complexity and process move analysis (inspired by descriptions of instrumental and framing design judgments; \cite{Murdoch-Kitt2020-sw,Gray2017-dx,Nelson2012-ov,Gray2015-qi}) to characterize the approaches that participants used to make the design of their action plan tractable. Across both of these forms of analysis, we relied upon a reflexive thematic analysis approach~\cite{Braun2021-dt}, acknowledging that our findings are impacted by our positionality as researchers and that our philosophical commitments and experiences as researchers of ethics shape how we formed interpretations of our co-creation data. 

In order to concretize our findings that described our participants' engagement in creating their action plans, we iteratively and collaboratively developed a set of roles and process moves that allowed us to distill the outcomes of our analysis of the artifacts and conversations the participants created during the co-creation sessions. The findings from our role and process move analysis are detailed in Section~\ref{sec:rq1}. In conducting our \textit{role analysis}, we began by considering the framework of Chivukula et al.~\cite{chivukula2021identity}, which characterized the ethical roles and identity claims that socio-technical practitioners embody when navigating ethical complexities. Through iterative analysis, we investigated the ethics-focused roles the participants in the co-creation sessions took on as they created their ethics-focused action plan, including evaluation of how these different roles were manifest, how those roles influenced the ways the participants navigated the session, and the kinds of action plan decisions that were motivated by these roles. Additionally, this analysis enabled us to describe how participants often took on different types of roles to navigate their felt ethical complexity depending on which part of their organizational ecology they were directing their action towards. We conducted this analysis by examining the types of problems and ethical concerns the participants raised, the method building blocks they selected to solve those issues, the goals they intend to achieve by solving those issues, and their reflections at the end of the session. The final set of roles from our analysis is detailed in Table~\ref{table:roles}.
In conducting our \textit{process move} analysis, we identified participants' use of framing, instrumental, and appreciative judgments~\cite{Nelson2012-ov} to describe how practitioners actively shaped their action plan and corresponding problem space. Through this analysis, we investigated the distinct judgments that participants in the co-creation sessions used to navigate their design space, including their management of felt ethical complexity, identification and iteration of problem scope, selection of relevant ecological components. These process moves reveal shifts in the participant's negotiation of their problem space and frame (e.g., relevant constraints, goals, items in or out of scope) and the appreciative judgment they used to inform values that were central, peripheral, or specifically excluded from their action plan. The final set of process moves from our analysis is detailed in Table~\ref{table:processmoves}. 

To more fully illustrate our findings, we selected three participants---representing diversity across professional role, types of roles and process moves utilized, and final action plans---to describe as cases. We use these case studies~\cite{Yin2009-vs} with a co-design participant as the unit of analysis to describe the interplay of participants' ethical design complexity through roles and process moves, and the relationship of these elements to the iterative design of the participant's action plan. These case studies are detailed in Section~\ref{sec:rq2}.

\begin{table*}[!htb]
    \begin{tabularx}{\textwidth}{p{.15\textwidth}X}
       \toprule
       \textbf{Process Moves} & \textbf{Through this move, \ldots} \\
       \midrule
      \textit{Refining/Contracting} & the practitioner narrows their design frame by identifying areas of focus or removing constraints, facilitating more focused attention to the detail of their action plan. This process move does not alter the overarching appreciative system used to evaluate the success of outcomes.\\

    \textit{Expanding} & the practitioner expands their design frame by including additional components, areas of focus, or areas of ecological complexity, facilitating broader coverage of functionality or use cases. This process move may completely change or alter the overarching appreciative system used to evaluate the success of outcomes.\\

    \textit{Diverging} & the practitioner alters the directionality of their design process based on emergent goals or interests, facilitating outcomes that appear more actionable or are better aligned with their goals. This process move is characterized by a change in the appreciative system that redefines what success means for the practitioner.\\

   \bottomrule 
     \end{tabularx}
     \caption{Process moves that participants used to navigate and shape their problem space.}
     \label{table:processmoves}
 \end{table*}

\subsection{Researcher Positionality}
The authors of this paper include researchers from two large, research-intensive public universities in the Midwestern USA and a university in India. The research team has previously engaged in multiple research projects relating to technology and design ethics, and as a group we have educational and professional training in design, psychology, ethics, and computing and are passionate about fostering ethical awareness in design and technology practice. We acknowledge that our research interests and professional objectives have motivated this study and informed our analysis process. We also acknowledge that our understanding of ethical complexity---as augmented by experiences of practitioners that we have identified in previous studies---impacted the form of the co-design materials that participants engaged with, and also shaped our facilitation practices through which we collected data.

\section{RQ\#1: Approaches to Structure Toolkit Design}
\label{sec:rq1}

In Section~\ref{sec:roles}, we will describe three primary roles that participants took on as they created their ethics-focused action plans, indicating how these roles enabled participants to activate their ethical focus and engage their felt ethical design complexity. In Section~\ref{sec:processmoves}, we will then describe three process moves that participants used to navigate the design of their action plans, including ways they managed the complexity of their ethical dilemma in relation to their appreciative (value-related) and instrumental (tool-related) judgments. 

\subsection{Roles}
\label{sec:roles}

\subsubsection{An Advocate}

An \textit{Advocate} represents instances where participants sought to take action based on their intrinsic interest and awareness of their ethical role within their team or organization. This role indicates the participant's interest in translating their personal awareness to others within their organization, and taking on this role facilitated participant's \textit{advocacy} for specific causes they felt would increase ethical engagement, including, for example: accessible design, privacy protection, design inclusivity, prioritizing user needs, encouraging open communication, or even the importance of sketching in the design process. The Advocate role often emerged early in the co-design session as participants considered what ethical dilemmas they faced and how they wanted to reconcile these dilemmas. For example, PS02C\footnote{Participants are referred to by an identifier throughout the findings section. PS indicates a practitioner session and SS indicates a student session, the number indicates which of the six total sessions the participant engaged in, and the final letter indicates the unique participant in that session.} (an Enterprise UX Research Lead) explicitly stated on their problem card ''\textit{I want to advocate for UX research or invest in UX research so that we can create user-centric products and services.}'' PS01B (an Enterprise UX Designer) linked their focus of advocacy towards a specific unethical phenomenon, framing their problem card around: ``\textit{interaction manipulation [\ldots] like dark patterns or nudges}.'' In addition to using a sensitizing concept to frame their potential advocacy, participants also employed the Advocate role to consider preempting future unethical events they felt should not be allowed to happen, or to forestall past events that they felt should not be allowed to reoccur in the future. For example, PS01A (an Enterprise Product Manager) described their concerns about the ethics of engaging with the metaverse: ''\textit{I've been really interested in the concept of the metaverse [\ldots] and it's exciting, but it also really scares me because I know that there needs to be a bunch of research and there needs to be a bunch of stakeholders looped in from the beginning to make sure that this is a technology that's used, you know, for the greater good and not for anything else. So I think I just have like a bunch of questions about it and I want to learn how to better be an advocate or put myself in a space where I can help advocate for like the better side of the technology than the negative.}'' Similarly, reflecting on their past industry experience in relation to ethical awareness recently acquired through their formal education, SS01A (a UX Design student) remarked: '' \textit{I'm in a Disability and Technoscience class right now and learning about technoableism and reflecting on my internship, I noticed that there were some things that should probably not happen in the future.}'' Participants that took on this role were often, in addition to playing the role of an advocate, open to taking actionable steps that would ensure that their action plans achieved the results they expect. For instance, SS01A started designing their action plan intending to communicate and advocate for ``accessible design'' within their organization. However, while creating their action plan, they realized that their advocacy would be more likely to thrive in an open-minded team, which prompted a brief exploration of the practicality of building such a team through a Reformer role before transitioning back to developing an advocacy-focused action plan to spread awareness of the need for accessible design within their organization.

\subsubsection{An Operationalizer}

An \textit{Operationalizer} represents instances where participants identified a component of their practice and experience that they felt related to ethical awareness, but did not situate that knowledge in relation to broader ecological complexity that may have given rise to or otherwise shaped the initial ethical concern. This role indicates the participant's interest in honing a small ``piece of the puzzle'' without addressing the ecological setting for their proposed action plan, either avoiding consideration of key stakeholders or otherwise limiting their treatment of ethical complexity. The Operationalizer role was the least common role participants took on during the co-design sessions and was more prominent among student participants as compared to practitioners. Operationalizers typically focused on their own professional role and responsibility, using this professional knowledge as a frame to explore how their action or in-action might impact, induce, or otherwise shape downstream unethical outcomes. However, unlike the Advocate or Reformer roles, participants embodying the Operationalizer role did not actively seek to define or engage with the complexity of those downstream unethical outcomes or the upstream forms of complexity that may shape the emergence of ethical concerns. For example, PS03F (an Enterprise Software Engineer) considered software bugs as a matter of ethical concern, but did not actively engage with the upstream ethical complexity that might have given rise to the software bug or the ethical impacts that might be produced downstream if bugs were left uncorrected. When describing why they felt software bugs were unethical, PS03F mentioned that \textit{``the thing that I've found in my career is that the second pass at something [\ldots] will always be like four times as efficient as the first time. And that's just, that's just how it works.''} Implicit in their sentiment is that software bugs arise due to a lack of due diligence and insufficient effort; hence, their overarching ethical frame that they used to operationalize their action plan was about ``\textit{chasing the constant dream of perfection---the perfect code,}'' thereby motivating them to design an action plan to enable them to eliminate errors in code production. SS02D, a student with industry experience as a UX Designer, also took on the Operationalizer role to navigate their ethical complexity while developing their action plan. In their case, they focused the design of their action plan with a goal of operationalizing and supporting their creativity and self-expression as a designer to mitigate ethical tensions, recognizing that ``\textit{advertisement is something that users hate---so to some extent, there's an ethical problem just before I do my design part}'' but desiring for their ``\textit{output to be valuable, both in terms of design creativity, as well as the value for users.}'' When their action plan was critiqued by another participant, they expanded the scope of their plan to include the implications of a lack of ethically-grounded creativity on end users. In general, participants taking on the Operationalizer role while creating their action plan possessed a more limited understanding and awareness of the nuances of their own ethical complexity, either framing professional values as ethical without describing the interplay of values (i.e., highlighting the efficiency of code without considering downstream negative impacts of buggy code to users or society) or identifying aspects of professional practice without considering the positive ecological impact of better support (i.e., using creativity not just as an indication of professional role but also as a tool to further interrogation of potential negative impacts of decisions using a speculative positioning).

\subsubsection{A Reformer}

A \textit{Reformer} represents instances where participants recognized that their vision of ethically-focused action was not equally shared by others in their organization---a situation they desired to change. This role indicates the participant's interest in building an action plan that would activate their intrinsic desire for ethical change in ways that might effect substantial change at the organizational or professional role level. This role was equally assumed by practitioners and students in the co-design sessions. Participants taking on this role often sought to change structures and processes within their organization that they deemed to be unethical, including changing their project scoping and approval process to ensure that potentially harmful projects are not approved and democratizing their design process to make it easy for any designer to utilize suitable design methods or processes. For example, PS03B (an Enterprise Data Scientist) stated on their problem card: ''\textit{I want to introduce an instant `stop project' criterion within our data project scoping process so that projects can be stopped when a potential harm is discovered and to ensure that projects are not launched until ethical release criterion are met,}'' thus demonstrating that this participant is not merely advocating for personal change, but want to induce and activate the change to reform their design team or organization. Participants taking on the Reformer role often began planning their desire for reform by identifying ways to advocate to and sensitize their organization or design team of either the need to change their existing processes to prevent an unethical event from occurring or to alter their process as a response to an ethical breach within their organization or team. For instance, PS01D (a UX Researcher) took on the Reformer role in order to activate their desired ethical change and align it with organizational values, taking into consideration the barriers they could encounter at the team or organizational levels such as design accountability and power dynamics in business. Similarly, SS02A---a student who came from a professional UX design background---started by taking on an Advocate role, seeking to build awareness and alignment in their team and organization to focus on user transparency. Through their action plan design process, they shifted towards the role of a Reformer by refocusing their efforts on defining responsibilities strongly within the organization and the coordination between these roles in ethical product delivery and design, which they felt would help to ``\textit{align design process with values of transparency,}'' ``\textit{sensitizing the team on why this would be beneficial to the users,}'' and ``\textit{aligning the value system mission of the company to the customers.}''
As participants using the Reformer role built out their action plan, they often transitioned back-and-forth from an Advocate role---which focused on modification of their own ethical practices---to a Reformer role that sought to create broader impact on organizational or disciplinary processes and structures, thereby making these more ethical practices the ``new normal'' and a shared goal within their organization or team. This shift between roles, and the kinds of action plan constraints represented, demonstrates that Reformers are usually interested in realizing material changes on the organizational or structural level and are not typically satisfied with only sensitizing the actors within their organization (including themselves) of the need to make those changes.

\subsection{Process Moves}
\label{sec:processmoves}

\subsubsection{Refining}

The process of \textit{refining} refers to the act of narrowing a design frame by identifying areas of focus or removing constraints, thereby facilitating more focused attention to specific kinds of detail in the participant's action plan. This process move does not alter the overarching appreciative system used to evaluate the success of outcomes, but rather focuses the participant's attention on scoping into more specific or constrained aspects of their original design space. Refinement was the most common process move amongst participants across all the sessions, and was used particularly often by practitioners. In multiple cases, participants used the refinement process to facilitate expansion to varying degrees, and a substantial number of participants focused almost entirely on refining their action plans without taking on explicit expansion or diverging process moves. Those taking on the ethical role of the Reformer more commonly refined their action plans in comparison to other process moves, likely because these participants were seeking to work within the footprint of what was already possible or available in their work context. The specific qualities of refinement also varied based on the stage of the workshop in which refining took place, where some participants began refining from the moment they articulated their goal on their problem card while others chose to refine only after a period of exploration and iteration on their action plan.
SS02A (a UX Design student) used the refining process move to %, who had experience as a practitioner in the past. Since the objectives of reformers are inherently on a larger scale (realignment and adoption on an organizational level), SS02A adapted to a direction that specifically 
target the alignment of team responsibility in service of their goal of improving transparent and honest design practices. While this participant primarily took on a Reformer role, they recognized the need for realignment and adoption on an organizational level, such as ``\textit{dividing the responsibilities, [such as] different stakeholders and their responsibilities}'' but focused their design efforts on team alignment while removing constraints relating to the organization at large. Thus, SS02A’s objective was to start small, with the latent assumption that the ``ripples'' of their action plan may later make larger ``waves'' on an organizational level to achieve their overall goal. %'' \textit{The other was how, in terms of as a whole, in terms of organization, like how organization can promote those ideas of transparency in their mission or value statements”}, which could be achieved through an action plan forged from continual refinement and focus. 
Another example of refinement occurred with PS03F (an Enterprise Software Engineer) whose goal was to optimize the process of bug fixing and assign ownership to issues, which they had framed as a matter of ethical concern. PS03F’s action plan focused on this specific process, and their efforts during the workshop was to hone their approach to encourage a more efficient experience: ``\textit{I realized halfway through creating it that it's very set in stone, like a bug triage plan. And because two, there's only really one way to make perfect code and that's to iterate on it and to find the issues, resolve them, learn from them and carry them into the future.}''  This judgment of what it meant to be more ``ethical'' for this participant was framed through the role of an Operationalizer, which when paired with their refinement focus, created a practical action plan that was useful in optimization but perhaps strayed away from typical views of what it meant to ``be ethical.'' Thus, in this case, using the refinement process move without considering other relevant details minimized---and perhaps even flattened---the participant's understanding of their ethical design complexity beyond the incomplete reification of ``optimization.'' 
Finally, as an instance of beginning their refinement later in the construction of their action plan, PS01B (an Enterprise UX Designer) employed the following approach: ``\textit{So enriching my scope, I thought about how to make my action plan more personal, thinking about the designer and making changes to do that. So I discarded the first step and completed the last one and put this one for what makes sense in this context, because last time I didn't.}'' In this case, PS01B used refining later in the workshop to cut out detail they had built in the initial round, using this constrained focus to further hone their action plan in a way they felt was more focused.

%Quotes

%\begin{itemize}
    %\item PS03F - Software Engineer - \textit{I realized halfway through creating it that it's like a very set in stone, like bug triage plan. And because two, there's only really one way to make perfect code and that's to iterate on it and to find the issues, resolve them, learn from them and carry them into the future.}
    
    %\item PS01B - UX Designer/Architect - \textit{So enriching my scope. I thought about how to make my action plan more personal, thinking about the designer and making changes to do that. So I discarded the first step and completed the last one and put this one for what makes sense in this context, because last time I didn't}
    %\item SS02A - UX Designer - SS02A adapted to a direction that specifically targeted team responsibility alignment within the larger space of designing transparently and honestly, \textit{“The other is dividing the responsibilities, like different stakeholders and their responsibilities.”} SS02A’s objective was to start small, and then use the ripples of their action plans to make larger waves on an organizational level to achieve his main goal, '' \textit{The other was how, in terms of as a whole, in terms of organization, like how organization can promote those ideas of transparency in their mission or value statements”}, which could be achieved through an action plan forged from continual refinement and focus.
%\end{itemize}

\subsubsection{Expanding}

The process of \textit{expanding} occurs when a practitioner extends their design frame by including additional components, areas of focus, or areas of ecological complexity, thereby facilitating or anticipating broader functionality, additional stakeholders, or more than one use case. This process move may completely change or alter the overarching appreciative system used to evaluate the success of outcomes, with the newly expanded set of constraints that define the new design frame indicating a prioritization of certain appreciative factors that may not have been present (or present to the same degree) in un-expanded form. The expanding process move was slightly more prevalent in students than practitioners, but appeared commonly in both participant populations. 
Participants utilized the expander process move in two primary ways, including: 1) creating an action plan which has an expanded focus as compared with their initial dilemma or goal, where they added elements they came to realize were salient to addressing their ultimate goal; and 2) shifting to this process role from the refiner process role to illustrate a potential macro- to micro-application of an action plan that had different forms of value at different organizational or structural levels.
In the first case, the expanding process move was typically undertaken when the participant felt that their in-development action plan required additional elements or focus areas to make it successful if it were to be applied within their chosen work context. For example, SS01B (a UX Design student) commenced the co-design session with the goal of understanding the ethicality of certain design decisions made by their organization. However, while building their action plan this participant expanded their area of focus to frame ethics in relation to the creative freedom they felt designers should have, since they believed that creativity could be a starting point for the organization to respect the users’ freedom and autonomy. In addition, SS01B knew they would need to involve additional stakeholders in relation to their goal, thereby expanding their field of action as well: ''\textit{I could relate my problem to trust and autonomy because if you're talking about users’ freedom of choices, then it's important to build trust between the organization and the users. I feel like it's not like you as a designer or you as someone cannot just take a decision, you have to convince the others and make them understand why you're doing what you're doing.}'' This expansion move involved not only a wider field of view with more actors in the organization, but also an appreciative frame that shifted from  a focus on ethicality in general to ethical considerations that could be guided by designer creativity.
%Participants utilized the expander process move in two primary ways, including: 1) creating an action plan which has an expanded focus as compared with their initial dilemma or goal, where they added elements they came to realize were salient to addressing their ultimate goal; and 2) shifting to this process role from the refiner process role to illustrate a potential macro- to micro-application of an action plan that had different forms of value at different organizational or structural levels. 
In the second case,
%This particular shift in process roles was more common in the student participants. For example, participant 
SS02C (a Product Management student) started the session with the goal to improve communication practices within their organization and relevant stakeholders so that concerns from each team are addressed and everyone would be aligned in the approach they were taking to address these concerns. To achieve this goal, the participant designed their action plan to map communication happening at various levels from macro to micro levels, including between stakeholders, among cross-functional teams, within the teams, and even in individual meetings. SS02C mentioned their desire for this expanded role of communication, noting the range of organizational facets they sought to re-shape: ''\textit{Group conflicts, our discussion of the worldviews and perspectives the different departments, how their approach was, what the issues were in their data collection---all of these things should be identified and also clarified by all the different stakeholders available, including the clients. [\ldots T]hen [we should have] a discussion about responsibility, what we take responsibility for, what each of the people are taking responsibility for, and creating a space which is non-judgmental and non-resistant, but open to discussing the issues. I think that is a culture within the organization rather than something a tool could fix.}'' PS01D, a UX Research practitioner shared an analogous example in their action plan, where they sought to ``\textit{get buy-in from different stakeholders and then how to collaborate across teams}''; as part of this goal, they expanded their focus to represent many different professional roles, including ``\textit{designers, PMs, and other stakeholders relevant to marketing}'' and used behavior and value cards from the shop to expand again when recognizing ``\textit{that organizations also have values. So maybe bringing those in to align people and then using those as a lens as well.}'' For this participant, they recognized more and more areas for potential expansion as their design process went on, reflecting: ``\textit{It’s about how with time we can slowly try to influence these different parts or where things would come in. I wish I had more time to actually go through everything and add in more things.}''

\subsubsection{Diverging}

The process of \textit{diverging} occurs when a participant alters the directionality of the their design process to facilitate outcomes that appear to be more actionable or which they feel are better aligned with their goals. This process move is characterized by a change in the appreciative system that redefines what success means for the practitioner. The shift in appreciative system can be either congruent with an existing appreciative frame with the addition of a new element that shifts its focus, or represent an entirely new appreciative frame that allows for new consideration of previously added action plan materials. The diverging process move was more common among student participants than practitioners.
For instance, SS02D came to the session hoping to design an action plan that would enable them to develop creative advertisements that did not manipulate users. However, during the session, they realized that this goal was too complicated to solve, and as a result, they diverged---shifting their goal to a different problem space: ``\textit{I started the session with the goal of developing creative design advertisements that are useful for users. However, maybe this question is too abstract to solve, and maybe I got confused about what kind of solutions I can build.}'' In this case, diverging resulted in a shift to a completely new design goal, recognizing that the initial goal was intractable or undesirable. However, the diverging process move also occurs when the participant realizes that a foundational problem needs to be solved before their action plan can become meaningful---often realized through a prior expansion process move. For instance, SS02B (a UX Design student) originally wanted to design an action plan to foster honest design practices and enhance user data protection practices. However, while developing their action plan, they realized that their objective might receive low uptake if the organization was not already sensitized to the need for user data protection. As a result of this realization, they diverged and expanded their scope to create awareness about the need for user data protection within their organization, thereby attending to both upstream and downstream considerations that framed their original appreciative focus. They articulated this divergent realization as follows: ''\textit{I started by wanting to understand how data security affects the user, particularly when their data is compromised. However, before we talk about this, we have to first of all as an organization discuss what kind of behavior would lead to a data security breach and what we can do as an organization to prevent such from happening.}'' Some participants also employed the diverging process move as a way of governing and maintaining control over ethical complexity. For instance, PS01A (an Enterprise Product Manager) remarked: ''\textit {As I was creating my action plan, I found it quite difficult because I kept on realizing that there were many, many more steps, and I was trying to figure out exactly where my plan would all fit in. And so I thought maybe I should start with an internal co-design where you go through and discuss project goals, ideas, technical restraints, among other things.}'' In this case, the participant recognized complexity through expansion process moves and then diverged in how they wanted their action plan to address their felt complexity---moving from an individually-focused action plan to one that had the potential to produce reform on the organizational level.

%Quotes

%\begin{itemize}
    %\item SS02D - \textit{I started the session with the goal of developing creative design advertisements that are useful for users, however, maybe this question is too abstract to solve, and maybe, I got confused about what kind of solutions I can build.}
    
    %\item SS02B - \textit{I started by wanting to understand how data security affects the user, particularly when their data is compromised. However, before we talk about this, we have to first of all as an organization discuss what kind of behavior would lead to a data security breach and what we can do as an organization to prevent such from happening.}
    %\item PS01A - \textit{As I was creating my action plan, I found it quite difficult because I kept on realizing that there were many, many more steps, and I was trying to figure out exactly where my plan would all fit-in ……  And so I thought maybe I should start with an internal co-design where you go through, and you know, discuss project goals, ideas, technical restraints, among other things.}
%\end{itemize}

\section{RQ\#2: Accounts of Ethical Complexity}
% How do participants engage the ethical complexity of their role as they created a bespoke ethics-focused action plan?
\label{sec:rq2}

\subsection{Praveen (PS03B): ``\textit{I want to change how we approve Data Science projects}''} 
Praveen is a Data Scientist in a B2B2C Enterprise company and has eight years of professional work experience. He expressed interested in fostering ethical awareness within his practice early in his workshop experience and mentioned that he is an avid follower of trends in Ethical AI and Ethics of Data Science. Praveen signed up to participate in the co-design session because he was concerned about the fairness of data-driven systems on protected communities and wanted to explore ways of mitigating its negative impact. He also mentioned concerns about the harmful effects of such systems on the data privacy rights of users. Across his action plan design process, Praveen's trajectory of engagement included increasing awareness of the ethical complexity of his specific work setting and the articulation of multiple strands of pragmatic action that could inform change on different levels of the organization. The appreciative system framing these goals sought to connect behavior(s) deemed to be unethical due to their negative or inequitable impacts with relevant actors and processes, and over time, Praveen was able to use his increased sensitivity towards this ecology to identify other ethical challenges he could address through a similar ecologically-grounded and multi-stranded approach. 

\textbf{Initial Goal.} As he began to reflect on ethical issues he had encountered, Praveen described instances where his company was ``\textit{tracking employee behavior to measure productivity,}'' indicating his past ``\textit{inability to find ways to balance fair lending with the profitability of a financial bank,}'' and his felt lack of ability ``\textit{to handle privacy concerns when working with DEI data.}'' In relation to his role as a data scientist, Praveen also expressed concerns about ``\textit{senior management overriding the decision of my team and not being transparent about why}'' and a lack of comfort regarding being viewed as the ``\textit{de facto fairness and ethics expert in the room.}'' Through his engagement with these aspects of his felt ethical complexity, Praveen centered his early goals on developing an ``\textit{instant-stop project criterion}'' that he could use to forestall the development of potentially privacy invasive or harmful systems. He mentioned potentially including ways to make it ``\textit{easier to align legal regulation with engineering practices}'' and monitor the performance of his company's datasets. %With respect to advancements in their field that he is concerned about, Praveen mentioned that he is concerned about the potential side effects of “facial recognition technology, automated gender recognition systems, automation of welfare and poverty services, Crypto/NFTs, and the Metaverse.” At the conclusion of the reflection sessions, when participants were asked to select one ethical concern from all the ethical issues they listed and to design an action plan to resolve them. Praveen at first wanted to focus on designing an action plan to resolve the issue of “companies tracking employee behavior to measure productivity.” However, after a brief reflection, he diverged his focus and instead selected to develop an ``instant stop project criterion.'' He provided a rationale for focusing on the instant stop criterion, mentioning that " 
He chose to focus on an instant-stop project criterion ``\textit{because when we talk about ethics and data science and ethics and AI, there's this concept of `don't build it, don't design it, don't deploy it.' If the technology is inherently harmful or if the project is found to be harmful after it is designed, then we should stop it.}'' Based on this goal, he wanted to design the action plan to get ``\textit{stakeholders to agree on `no-go' conditions before a project is started}'' and also ensure that his organization's ``\textit{project release criteria include [examination of] fairness and privacy concerns}'' before a project is approved. Praveen's ultimate objective was to ensure that ``\textit{projects can be stopped when a potential harm is discovered}'' and that ``\textit{projects are not launched until ethical release criteria [for privacy and fairness concerns] are met.}'' Before setting out to design his action plan, Praveen had an implicit awareness that he had the power to alter the project implementation process within his organization since he was already acknowledged as an algorithmic fairness expert within his organization. This acknowledgement of his ethical authority within the organization---shared by both Praveen and members of his organization---provided a grounding for his action plan design, through which he primarily assumed the Reformer role to reshape the ethical posture of his organization.

\begin{figure*}[!htb]
    \centering
    \includegraphics[width=\textwidth]{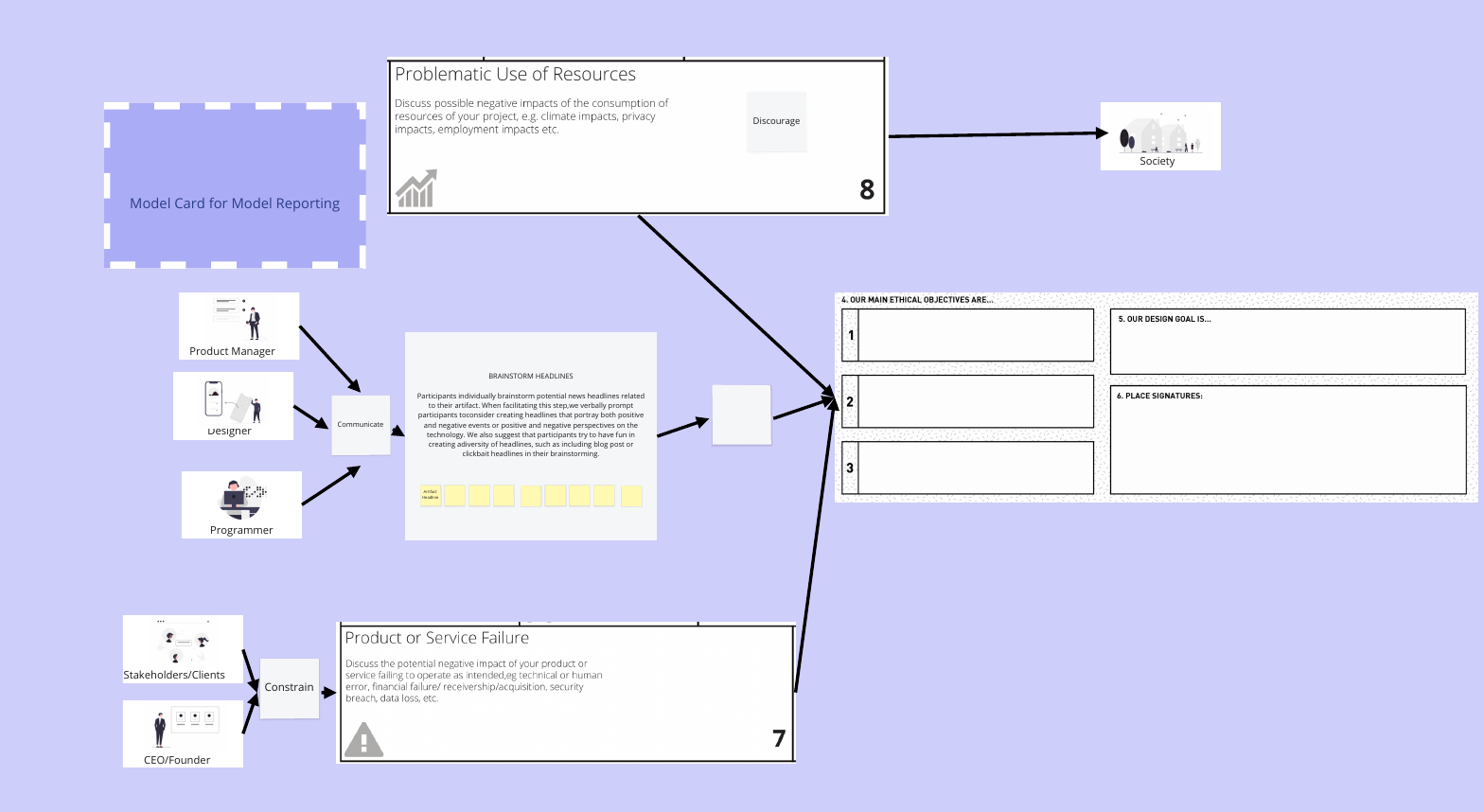}
    \caption{Praveen's final action plan.}
    \label{fig:praveen}
    \Description[Praveen's final action plan]{}
\end{figure*}

\textbf{The Shop.} To support his goals and initial framing of ethical concerns, Praveen selected four ethics-focused method building blocks, including: brainstorming headlines; determining the main ethical objectives of their organization; discussing the possible negative impacts of user privacy violation to their project; and discussing the potential negative impact of their product on users if it fails to function as intended. Praveen also brought in his own method block, which was titled ``model cards.'' He selected a range of actors he felt were relevant to the implementation of his action plan, including: product manager, designer, programmer, stakeholders/client, and CEO/Founder. %To activate his goals, he sought to constrain, communicate, and evoke certain values amongst the actors that he had identified.
 
\textbf{Initial Design.} While designing his action plan (Figure~\ref{fig:praveen}), Praveen employed multiple ethical roles and process moves simultaneously based on the aspect of his organizational and ethical ecology he was targeting at that point in his action plan. His action plan contained three parallel layers---representing expanding process moves---that eventually converged to inform positive outcomes for Praveen's organization. In the \textit{first layer} (Figure~\ref{fig:praveen}, bottom), Praveen contracted his problem space, allowing him to focus on Reforming his organization's project approval process, refining his goals to specifically constrain the organization's CEO/Founder, Stakeholders, and Clients from approving projects that could be harmful to users. He achieved this goal by Reforming how the decision makers approve projects to include a layer that sensitizes the group towards the potential negative impact of a harmful project on their organization. In the \textit{second layer} (Figure~\ref{fig:praveen}, middle), Praveen used a combination of the Advocacy and Reformer roles, contracting his problem space to focus on sensitizing his teammates towards the importance of designing ethical products, Reforming how the team determines that projects are ready to be deployed. He described his design intentions as follows: ``\textit{these are exercises the team goes through to envision the potential ways that the product or tool could adversely impact people or society at large. And after coming up with those different hypothetical doomsday cases, then we agree on what are the ethical objectives of the project? We'll write them down and then write our goal for the project. And then everyone signs off and says these are what we agreed are the key values---ethical values of the project.}'' In the \textit{third layer} (Figure~\ref{fig:praveen}, top), Praveen used a combination of the Operationalizer and Reformer roles while contracting his problem space to focus on examining the potential side effects of problematic use of data resources such as model training on the environment and the cascading impact of environmental degradation on society at large. After designing the first draft of his action plan, Praveen commented that he was concerned about the practicality of implementing this plan in real-life professional and project settings---based on an implicit expansion when viewing all three layers together and a combination of Reformer strategies that he did not have full control over.

\textbf{Test Drive.} In the Test Drive room, Praveen diverged and reframed his problem space in order to explore how well the first draft of his action plan could support him in resolving ethical issues relating to privacy concerns when working with DEI data. His selection of this specific issue was based on his prior experience with the sensitive data and the trade-offs that emerge in its use: ``\textit{The ethical dilemma I chose was this project I had historically worked on, which is can we look at employee data to promote DEI efforts? I think there was a very interesting trade-off there and that promoting the diversity equity and inclusion is great. But when you're looking at employee HR data, that's very sensitive information. So, this is a trade-off between the privacy of your employee’s data versus the potential societal good or organizational good that can come from improving being able to optimize diversity inclusion efforts.}'' Through engagement with the action plan and this new goal, Praveen found the action plan to map well to this new concern: ``\textit{I think it really does translate well in the sense that, instead of let's say brainstorming headlines, it would be more so brainstorming emails or negative internal reports that could happen as a result of this type of DEI analysis. [\ldots] You know, maybe some senior managers are happy that [\ldots] we have been able to improve employee retention for this particular demographic group of say Black and Hispanic men by 25\%. But then there's also the also harmful emails of like, oh, hey, you've accidentally leaked people's private information through this analysis.}'' Following this reflection, Praveen further refined his action plan to make more prominent the goals that he was trying to achieve at each stage of the action plan. For example, instead of brainstorming on headlines, he added sticky notes to show that this portion of the action plan could also include brainstorming on emails, social media messages, and other ways in which people could react to the events that are reported out from their organization.

\subsection{Alex (PS03F): ``\textit{My code needs to be perfect}''}

Alex is a Software Engineer in a B2B2C Enterprise company and has five years of professional work experience working. He is aware of calls for software developers to be more ethical and is interested in equipping himself and gaining capacity to support his own ethical awareness. Alex signed up to participate in the co-design session because he was interested in exploring ways to introduce ethics in his practice as a software engineer and to explore how to utilize ethics to make his organization's software development process more ethical by considering the implications of imperfect code. Across his action plan design process, Alex's trajectory of engagement included a largely circular operationalization of ethics in relation to code quality that never resulted in broad ecological awareness of how issues such as code quality might produce negative impacts, or how it may be impossible to achieve perfection within a particular professional role. The appreciative system framing these goals was focused on technical quality and did not grow to account for other appreciative systems that could capture other forms of ethical impact in relation to software engineering practices.

\textbf{Initial Goal.} As Alex began to reflect on ethical dilemmas they had faced, Alex initially struggled to identify what would represent an ``ethical encounter.'' However, after a brief reflection, he identified a range of issues that he had encountered in his professional role relating to code quality, contributions, and compensation, such as: ``\textit{knowing that a bug exists in their software but it hasn’t surfaced yet; knowing the salary discrepancies between members of staff; uneven group contributions between members of their development team; and an unethical payroll software.}'' Alex quickly began to converge his efforts on code quality, describing his discomfort when finding ``\textit{imperfect code that may perform badly in live environments, inaccurate reporting as a result of bad code, [and] upset clients demanding answers and solutions}.'' %When he was asked about the things he wished to do but could not, Alex mentioned that he wished he could take more time off from work. Concerning advancements in their field that he is concerned about, Alex mentioned that he is concerned about automation taking over the jobs of people. He also mentioned that he is concerned about Artificial Intelligence systems and media over-saturation. After the ethics-focused reflection activity, Alex decided to focus the design of his action plan around resolving what he characterized as an ethical concern, “imperfect code”. The context for his ethical concern was during code reviews or client meetings. 
His objective for focusing on imperfect code in the context of code reviews or client meetings was to prevent someone else from having to spend time to go back and fix the code, supported by an analogy Alex made between low marks on assignments in school and his workplace experience: %. Providing more rationale for electing to focus on this plan, Alex mentioned that  ''\textit{the best way of describing it is that if you do poorly on an assignment in school, you get a lower mark. 
``\textit{If you do poorly on something in the workplace [\ldots] your contribution has to be a hundred percent. And if it's anything less than perfect and it's not good enough.}'' This choice of ethical concern and rationale provided by Alex is a manifestation of his level of both ethical awareness and what he considers as his ethical complexity. Building on this issue of code quality, Alex took on an Operationalizer role and further contracted his problem space to refine his concerns. 

\begin{figure*}[!htb]
    \centering
    \includegraphics[width=\textwidth]{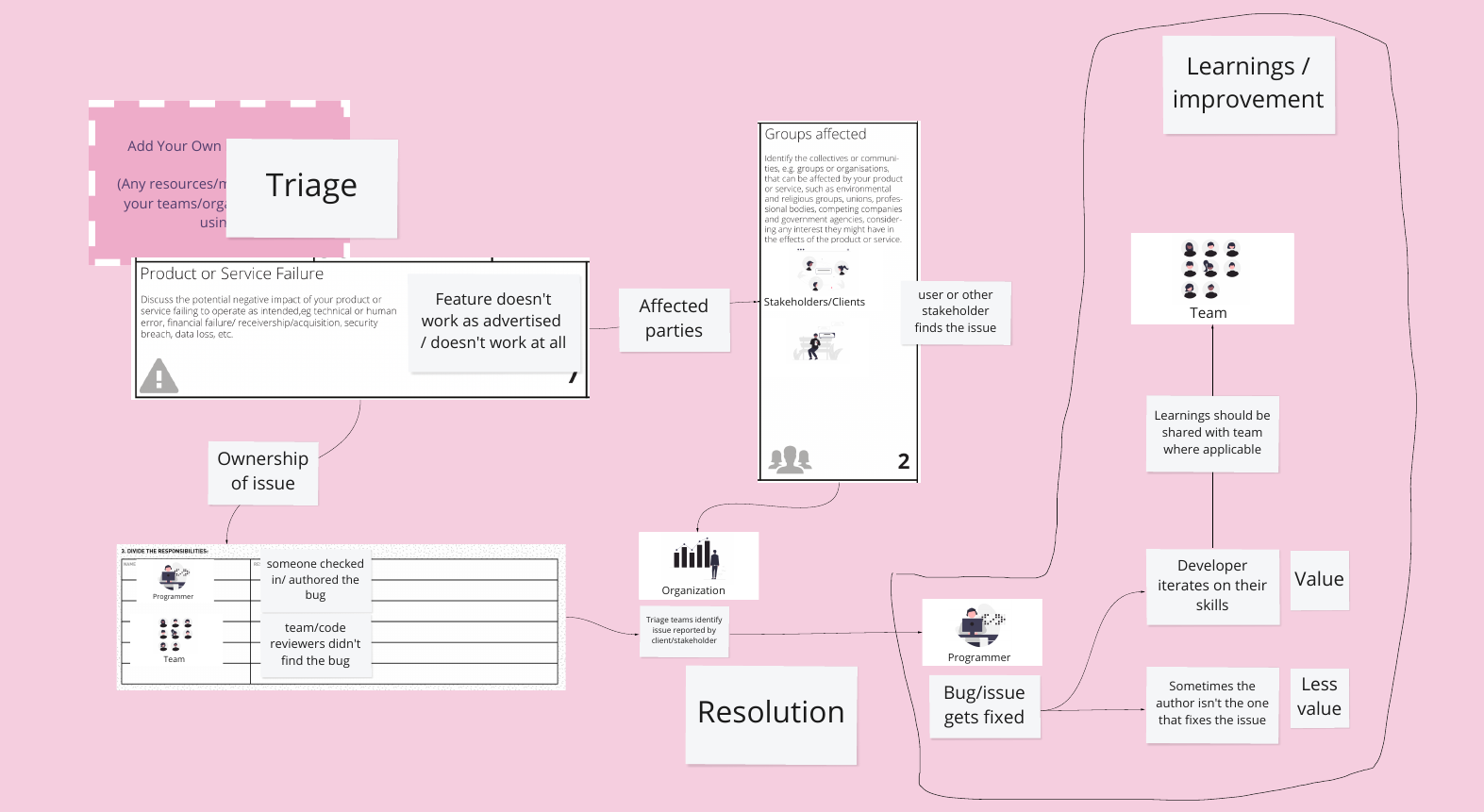}
    \caption{Alex's final action plan.}
    \label{fig:alex}
    \Description[Alex's final action plan]{}
\end{figure*}

\textbf{The Shop.} While shopping for building blocks to help him design his action plan, Alex initially expanded his problem space to explore how his code could impact users, using this awareness to select three building blocks, including: determining groups, communities, or organizations that will be affected if their product fails; exploring how his team would react if their product or service failed to function as intended; and dividing responsibility among team members to identify responsibility for each task. He also identified various actors that would play a part in his action plan, including the programmers responsible for building the product, the team of the programmers, and the organization as a whole.

\textbf{Initial Design.} While designing his action plan (Figure~\ref{fig:alex}), Alex exclusively employed the Operationalizer role to form an action plan that identified how individual programmers, his team, and the entire organization would react to the cascading consequences that arise from a software bug. Specifically, his action plan contained three layers that included the Triaging layer, the Resolution layer, and the Learning/Improvements layer. Across these three layers, Alex navigated the problem space through the refiner process move, working entirely within his existing awareness and framing of code quality and only addressing other stakeholders or consequences through this lens. In the \textit{Triaging} layer (Figure~\ref{fig:alex}, left), Alex explored ways of responding to product or service failures that arose due to defective code. The triaging activities included identifying features that did not work as intended, the parties or groups that the issue affected, and the software engineers that were responsible for authoring and reviewing the defective code. The \textit{Resolution} layer (Figure~\ref{fig:alex}, center) identified how the programmers and their team responded to and fixed the issues identified by the client, with the goal of problem resolution. In the \textit{Learning/Improvement} layer (Figure~\ref{fig:alex}, right), Alex envisaged that the team responsible for the software bug could share what they learned so that the entire organization could learn from their experience. Across these three layers, Alex's goal was to guarantee perfection within his team and organization, framing ethical support as operationalized solely through code quality: ``\textit{everything has to be perfect all the time. Nothing can be not perfect because if it's not perfect, then it doesn't meet the spec.}''

\textbf{Test Drive.} In the Test Drive room, Alex further refined his problem space, selected an additional focus on inaccurate reporting and upset clients demanding answers/solutions, which was framed as an ethical concern. Continuing with an Operationalizer role, Alex felt that ``\textit{both concerns [poor code and inaccurate reporting] are bugs in their own way}'' and could benefit from the same process flow that he had designed in his initial action plan draft. This operational quality of the final plan was identified by Alex ``\textit{very rigid, but that’s the nature of the beast,}'' alluding to the idea that the software bug resolution process needs to be rigidly governed before it can yield what he felt were ethically-positive outcomes.

\subsection{Sarah (SS01A): ``\textit{I want to create a safe space to talk about ethical issues}''}

%\paragraph{\textbf{Background}}
Sarah is an undergraduate student at a large, research-intensive, public university in the Midwest, USA. Her undergraduate major focuses on Human-Centered Computing, and although she is still a student, Sarah had recently completed a three-month internship working as a UX Designer in a B2B company. Based on her educational and professional experiences, she believes that she can use design to make a positive impact on society, and decided to participate in the co-design session in order to explore more ways to empowering herself to promote ethical outcomes when designing products.  Across her action plan design process, Sarah's trajectory of engagement included the iterative identification of multiple strategies to build awareness of values in design teams, using a consistent overall focus on communication regarding ethical issues that manifest in different groupings of stakeholders and activities to support her overall goals. While her draft action plan was perhaps overly ambitious, her identification of multiple forms of pragmatic action---particularly in its refined form after the test drive---could inform increased communication about values with specific combinations of stakeholders.

%\paragraph{\textbf{Ethical Reflections and Problem Card}}

\textbf{Initial Goal.} As Sarah reflected upon some of the ethical dilemmas that she had encountered in her internship, she identified instances of misalignment where ``\textit{my personal political opinions did not align with the company, my personal beliefs did not align with some coworkers, and I could not have a conversation about 'touchy topics'}.'' %She also mentioned that she had a dilemma “participating in emotionally charged team meetings make it hard to say what really needed to be said” When asked about situations that made her feel uncomfortable Sarah mentioned that 
She also felt uncomfortable with ``\textit{company approaches to UX [that] were vastly different than mine}'' and that she ``\textit{felt apprehensive to bring it up as I did not want to upset or offend anyone}.'' %She also mentioned that she felt uncomfortable that her company is still using the design thinking innovation methodology and that the opaque culture of the company made it difficult for people to be forthright during team communications. When asked about the things she wish to do but could not, Sarah wished she could “engage in more of the UX workshops that were held for interns, explore different types of projects in the industry - was handed something (I had little interest in) and told to go with the flow, work on something more tangible”. With regards to some advancements in technology that she was concerned about, Sarah mentioned she is concerned about “AI systems that facilitate systemic issues from our physical world to our online world and consequences of harmful algorithms that have created society-wide issues” 
Building on these concerns in conjunction with some of her academic coursework, Sarah began to position her goals for the co-design session in relation to \textit{technoableism}: that is, the idea that technology can be used to solve all human problems. Using this concept as a frame, Sarah sought to design an action plan that ``\textit{makes others aware of possible accessibility issues/needs and to create safe meeting spaces for everyone to speak to their thoughts and concerns},'' enabling ``\textit{designers to create more accessible and less ableist design and to allow for others with disabilities to work in safer spaces/teams}.'' This ultimate objective was strongly motivated by Sarah's reflection on her internship through the lens of a course she was enrolled in at the time of the co-design workshop: %Providing a rationale for why she selected this action plan, Sarah mentioned that 
``\textit{I'm in a Disability and Technoscience class right now, and learning about technoableism and reflecting on my internship, I noticed that there were some things that should probably not happen in the future [\ldots] in team meetings specifically. So that's what I wanted to work on. And it's just about creating a safer and more open space to have discussions about design versus just listening to what the most dominant person is saying because they don't allow anyone else to talk}.'' % And this team was interdisciplinary. So, everyone was coming in with different backgrounds and not everyone was aware of like accessibility issues and not to speak to their opinions and what they actually care about.” 
Implicit in Sarah’s goal is the assertion that reforming her team's meeting structure would make it possible for her and other team members to better advocate for values such as accessibility and inclusivity. Sarah addressed this goal through the dual roles of an Advocate and a Reformer while creating her action plan, where she sought to reform her team meeting structure so that she would be able to Advocate for values such as accessibility during these meetings.

\begin{figure*}[!htb]
    \centering
    \includegraphics[width=\textwidth]{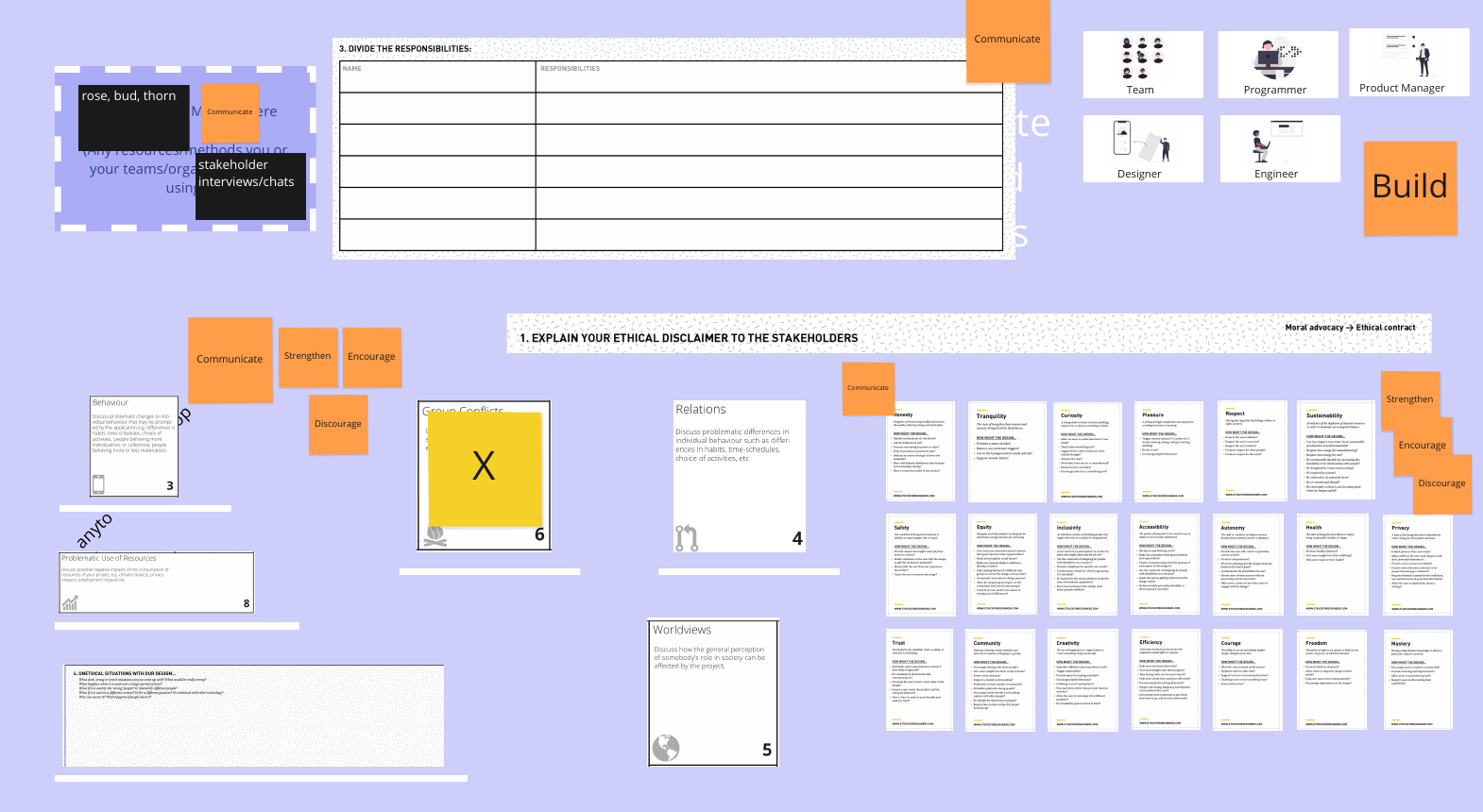}
    \caption{Sarah's final action plan.}
    \label{fig:sarah}
    \Description[Sarah's final action plan]{}
\end{figure*}

%\paragraph{\textbf{Shopping for Ethics-focused methods and designing their action plan}}

\textbf{The Shop.} To support her action plan goals, Sarah selected a range of ethics-focused method building blocks that included: resolving team conflicts, managing team relations, dividing ethical responsibilities between members of the team, and discussing potential unethical outcomes from their products. She also selected building blocks that enabled her to consider how to prevent negative impacts on users and the environment and identifying positive values to guide her work, including privacy, autonomy, respect, trust, inclusivity, and accessibility, among others. Sarah selected a range of potential actors that would be present on her project team, including a product manager, programmer, engineer, designer, and the team as a whole. Sarah also added her own methods, including: ``rose, bud, and thorn'' to help her explore the positive and negative dimensions of a project, and ``stakeholder interviews and discussions'' as a means of sensitizing decision makers. In explaining her rationale, Sarah sought to identify ``\textit{activities and methods that would facilitate conversations [\ldots] So dividing responsibilities [includes] questions to reconsider during the process of the project that they're working on. [\ldots] %And then for people, I put basically the whole team and 
I also added stakeholder interviews and chats because they're not going to be fully involved in the process, but [\ldots] they still need to be in the loop.}''% as well as rose, bud, thorn, because that's a good team meeting facilitator.”
 
\textbf{Initial Design.} While designing her action plan (Figure~\ref{fig:sarah}), Sarah fully adopted the dual role of an Advocate and Reformer and selected verbs that articulated her team and communication focus, including: encourage, discourage, strengthen, and build. She activated these verbs in different segments of her action plan, depending on the values she wanted to foreground or forestall within their organization. Specifically, her action plan focused on two groups within her organization: decision-making actors and technology-focused actors. For the \textit{decision-making actors}, Sarah adopted an Advocate role and utilized the ``rose, bud, and thorn'' method to communicate the importance of adopting ethical decisions (Figure~\ref{fig:sarah}, top left). Building on the awareness that could be gained through this portion of her action plan, Sarah used an ethical disclaimer building block (Figure~\ref{fig:sarah}, center right) to encourage communication around ethical issues, thereby creating a space of Advocacy for moral values such as privacy, respect, and autonomy. For the \textit{technology-focused actors}, Sarah used a building block to divide the ethical responsibilities (Figure~\ref{fig:sarah}, top center), suggesting new patterns of communication as a Reformer that team members could use to embody positive values such as privacy and trust, while discouraging team members from adopting design practices that might negatively affect the users and the environment.

%\paragraph{\textbf{Test Drive \& Reflection}}
\textbf{Test Drive.} In the Test Drive room, Sarah considered how her draft plan could enable her to address issues of company propaganda, building on her felt misalignment of political opinions that she recognized in the goal-setting stage. While Sarah's initial draft was quite complex, addressing a wide swath of organizational complexity, in describing the performance of her action plan in this new context, she reflected: ``\textit{I think a lot of things fail in [the company-wide meetings] context because it's kind of hard to talk and express your true feelings about certain things that are being implemented in the company when there's like 200 people on the call.}'' Following this realization, Sarah sought to refine her action plan, resisting some of her previous expansion and instead contracting her problem space to address only specific actors within the organization instead of seeking to focus on the entire organization. Through this realization she shifted from a broad Reformer focus that sought to shift numerous organizational communication practices and instead address specific communication practices through the Advocate role in contexts that were more closely under her control.

\section{Discussion}

\subsection{Challenges Practitioners Faced when Building Bespoke Action Plans}

Building on our findings in this study, we seek to better understand what our participants struggled with and how these method design processes might be better scaffolded. Of course, the roles and process moves that we have identified are---at some level---obvious. Design and technology practitioners and students seeking to make ethical changes must confront the locus of the change they seek to promote, with those already comfortable with their own ethical positioning often desiring to \textit{reform} their organization or profession and those with less experience interrogating their ethical role seeking to \textit{advocate} for practices closer to their own experiences and practices. Those that have already pre-framed their ethical concerns and areas for potential support may primarily \textit{operationalize} their current knowledge of the situation when considering what kinds of impact of potential change they seek to support. Similarly, as with any process or organizational change, a designer must consider which types and numbers of constraints allow the situation to feel tractable and malleable. Design practitioners and students, depending on their knowledge of their ecological setting, may easily recognize areas to ``scale up'' their action plans through \textit{expansion} or \textit{diverge} from their original goals after recognizing new aspects of ecological complexity through reflection. Similar to practitioners workers as \textit{operationalizers}, practitioners that are most confident in their knowledge of their existing ecological complexity may focus their efforts primarily on \textit{refinement}, with the assumption that they have already identified root causes and practices that need support.
 
However, these trajectories of engagement with ethics-focused action plans revealed an important interplay between felt ethical complexity and the use of method building blocks to form and iterate upon an action plan. Some practitioners struggled to break out of the box of their own professional role, operationalizing ethics in relatively narrow ways that negatively impacted their ability to have broader impact. Others recognized \textit{too many} ecological links between their own role and the organization or industry at large, and became bogged down in trying to fix everything. This causes us to question: \textit{How can we moderate or mediate these relationships to find a tractable frame more quickly?} and \textit{What form(s) of knowledge and scaffolds are necessary, desirable, or ideal to instigate and support iteration of an ethics-focused action plan?}

First, the use of existing knowledge---both through the provision of building blocks and common ecological elements and verbs---was overwhelming for many participants, often because they had no strong mental model for what a final toolkit or action plan might ``look like'' or at what level it might be used to operationalize or shift ethical focus in their organization or professional role. \textit{How can we provide frames for different levels of engagement, perhaps pre-framing the design space in relation to time (e.g., things that could change this week versus over a year), type of interaction (e.g., convincing a manager, building team alignment), or combination of stakeholders (e.g., something done alone versus with members of two or more disciplinary or professional roles)?}

Second, participants generally were able to identify \textit{many} goals they wanted to address, and multiple levels of complexity that could be considered alongside these numerous potential goals. Thus diverging and expanding activities were---for many participants---a means of deciding what impact they wanted to have during the workshop, and may indicate a need to support the creation of many \textit{different} action plans---representing different purposes, scale(s) of desired change, audiences, and use cases. \textit{How can we scaffold the creation of action plans as an everyday activity, and not just one that you need to complete a single time? How can we motivate the right scale of action plan so it actually gets used, recognizing that all action plans will also be iterated on while in use?}

Third, while we did not track participants' use or implementation of their action plans in their everyday work context, numerous instances in the action plan design process indicated their consideration or ``projected use'' of the action plan as one trigger for iteration or refinement. Future versions of co-design scaffolds to support action plans might include intentional periods of priming, implementation, incubation, and iteration over a period of time---perhaps weeks or months---to better map the intentions and goals of practitioners with the realities of their practice, shifting the action plan from ``just another method'' to a meaningful extension of one's praxis. \textit{How can we express action plan development and iteration over time in relation to dynamic work practices, building resonance through reflection and reflexivity?}

\subsection{Leveraging and Supporting Instrumental Judgments to Enable Practitioner Engagement in Ethics-Focused Work}

Building upon prior work that has described how method designers utilize knowledge and a range of creative constraints to make the design space for a new method tractable, we are able to identify how our co-design scaffolds functioned in enabling design and technology practitioners to build their own bespoke action plan, which we frame here as a bespoke \textit{design method}. We build upon two primary categories of decisive constraints---proposed by Biskjaer and Halskov~\cite{Biskjaer2014-gy} and operationalized for method design practices by Gray et al.~\cite{Gray2022-kv}---\textit{intrinsic} and \textit{self-imposed} constraints (including sub-types referenced below) to map the participants' use of creative constraints in structuring their action plan design process. 

\textit{Intrinsic constraints} framed participants' engagement in the workshop, including their understanding of their work environment and beliefs about  how methods might be used as a type of knowledge to support their work practices. First, constraints related to the participants' \textit{Epistemological Framing} surfaced in relation to their initial desire to participate in the workshop (a form of self-selection bias in its own right), including the goals and motivations they brought, relevant knowledge they had about their felt ethical design complexity based on previous industry and educational experiences, and their pre-conceived notions about what was or was not ``ethical'' in relation to these practices. Second, constraints related to the participants' \textit{Pragmatic Activation} linked their goals for engaging in or building methods supports for their practice, including their focus and desired outcomes stated on their problem card, with connections to the complexity of their work context, answering the question: What can methods accomplish to better support \textit{my} practice?
    
\textit{Self-imposed constraints} were identified by participants as salient and intentionally applied to shape their design space. We recognized the interplay of three different types of self-imposed constraints that impacted participants' design of their action plan. First, constraints relating to the \textit{Identification of Methodological Insufficiency} impacted the structure and purpose that participants set out for their action plan, including the ethical dilemma or problem they selected and the kinds of conditions they set out to change or re-shape through the introduction of a toolkit. This selection was primarily realized through the Advocate, Operationalizer, and Reformer roles. Second, constraints relating to \textit{Selection of Opportunities within the Design Ecology} included the articulation of embedded assumptions about what their discipline or professional role could contribute, and how this role could relate to other members of the organization or the organization at large. This selection was operationalized through the Expander, Refiner, and Diverger process moves. Third, constraints relating to \textit{Framing through Prior Design Knowledge and Intention} structured the design of the action plan, including method ``building block'' elements that we provided that appeared salient to the participants, their selection of other potential actors and verbs in the DIY room, and the visuospatial organization of these elements in their final action plan.

Overall, we found that participants were able to relatively readily identify method building blocks to support their toolkit design work. This implies new opportunities to disseminate, categorize, and make discoverable not only a range of methods or toolkits, but also to ``atomize'' these forms of design knowledge in ways that support re-use, re-organization, and the generation of completely new approaches to supporting ethically-focused work. For instance, in building on the metaphor provided by the \textit{Atomic Design} methodology~\cite{Frost2016-jc}, we might identify existing toolkits and methodologies such as Microsoft's Inclusive Design toolkit or the overarching methodology of Value Sensitive Design as ``organisms'' which implicitly contain many ``molecules'' that enable the functioning of the overall system. We ask, building on our co-design engagement with practitioners, how might these ``molecules'' become more directly parseable and tractable as design objects, and further---how might we then break down these ``molecules'' that often represent distinct design methods further into ``atoms'' that contain individual building blocks that might enable new downstream configurations of methods.

\section{Implications and Future Work}

While existing method design and implementation practices have focused attention on the method prescription itself, our work identifies new opportunities to consider practitioners themselves as best-placed to engage in method design. Rather than viewing method design as distant from practice and revealed primarily or only through method prescriptions, we question what opportunities could be realized for scholars, educators, and practitioners alike if we view the majority of practitioners as \textit{capable of creating tools to support their own work practices}. This reorientation of method design practices could draw on histories of tool use and adaptation in other creative contexts, such as the creation of ad hoc tools in hackerspaces~\cite{Bardzell2014-cx} or the formation of customized work practices, software, and collaborative techniques to support the creation of fan art~\cite{Gross2013-qf,Allred2021-fz}.
We have also framed the need for additional types of design knowledge---including method building blocks (i.e., ``atoms''), methods (i.e., ``molecules''), and toolkits (i.e., ``organisms'')---to support designerly efforts that are conducted by individual practitioners, design teams, and organizations. While method prescriptions have become increasingly standardized in some ways over the past decade, drawing on both the success of IDEO's Design Thinking framework and the popular \textit{Universal Methods of Design} text, the creation of a wholly new collection of methods (since none of the ethics-focused building blocks we used in this study are present in either existing collection of methods) offers the opportunity to question how---and in what presentation formats---methodological guidance to support ethically-centered practice could be structured. Future work could include the creation of scaffolds and other supports to aid practitioners in identifying salient components of methods, at a number of levels, that could form or inform bespoke practice-resonant methods. Additionally, scholars and educators could investigate how and at what levels of fidelity these bespoke methods should be specified to support differing types of performance.

\section{Conclusion}

In this paper, we report on trajectories of toolkit design undertaken by a range of technology and design practitioners, revealing patterns of support that were useful for these practitioners in building toolkits that were resonant with their practice and also opportunities to better situate and support the creation of bespoke design knowledge that has ecological resonance for practitioners within their organizations. We identified three roles that practitioners took on in developing their toolkit---including advocacy, operationalization, and reformer-focused attitudes towards their toolkit and its implementation. We also identified three different process moves that practitioners used to engage with the framing of their toolkit, including refining, expanding, and diverging moves that enabled or constrained their ability to address the felt complexity of their ecological setting. We conclude with opportunities for these method design efforts to be better scaffolded, and call for new ways to categorize and organize design knowledge to support ethically-focused design and technology practices.

%%
%% The acknowledgments section is defined using the "acks" environment
%% (and NOT an unnumbered section). This ensures the proper
%% identification of the section in the article metadata, and the
%% consistent spelling of the heading.
\begin{acks}
This work is funded in part by the National Science Foundation under Grant No. 1909714. 
\end{acks}

%%
%% The next two lines define the bibliography style to be used, and
%% the bibliography file.
\bibliographystyle{ACM-Reference-Format}
\bibliography{trajectories}

%%% -*-BibTeX-*-
%%% Do NOT edit. File created by BibTeX with style
%%% ACM-Reference-Format-Journals [18-Jan-2012].

\begin{thebibliography}{59}

%%% ====================================================================
%%% NOTE TO THE USER: you can override these defaults by providing
%%% customized versions of any of these macros before the \bibliography
%%% command.  Each of them MUST provide its own final punctuation,
%%% except for \shownote{}, \showDOI{}, and \showURL{}.  The latter two
%%% do not use final punctuation, in order to avoid confusing it with
%%% the Web address.
%%%
%%% To suppress output of a particular field, define its macro to expand
%%% to an empty string, or better, \unskip, like this:
%%%
%%% \newcommand{\showDOI}[1]{\unskip}   % LaTeX syntax
%%%
%%% \def \showDOI #1{\unskip}           % plain TeX syntax
%%%
%%% ====================================================================

\ifx \showCODEN    \undefined \def \showCODEN     #1{\unskip}     \fi
\ifx \showDOI      \undefined \def \showDOI       #1{#1}\fi
\ifx \showISBNx    \undefined \def \showISBNx     #1{\unskip}     \fi
\ifx \showISBNxiii \undefined \def \showISBNxiii  #1{\unskip}     \fi
\ifx \showISSN     \undefined \def \showISSN      #1{\unskip}     \fi
\ifx \showLCCN     \undefined \def \showLCCN      #1{\unskip}     \fi
\ifx \shownote     \undefined \def \shownote      #1{#1}          \fi
\ifx \showarticletitle \undefined \def \showarticletitle #1{#1}   \fi
\ifx \showURL      \undefined \def \showURL       {\relax}        \fi
% The following commands are used for tagged output and should be
% invisible to TeX
\providecommand\bibfield[2]{#2}
\providecommand\bibinfo[2]{#2}
\providecommand\natexlab[1]{#1}
\providecommand\showeprint[2][]{arXiv:#2}

\bibitem[Allred and Gray(2021)]%
        {Allred2021-fz}
\bibfield{author}{\bibinfo{person}{Alyse~Marie Allred} {and}
  \bibinfo{person}{Colin~M Gray}.} \bibinfo{year}{2021}\natexlab{}.
\newblock \showarticletitle{``Be Gay, Do Crimes'': The {Co-Production} and
  Activist Potential of Contemporary Fanzines}.
\newblock \bibinfo{journal}{\emph{Proceedings of the ACM on Human-Computer
  Interaction}} \bibinfo{volume}{5}, \bibinfo{number}{CSCW2}
  (\bibinfo{year}{2021}), \bibinfo{pages}{Article 376}.
\newblock
\urldef\tempurl%
\url{https://doi.org/10.1145/3479520}
\showDOI{\tempurl}


\bibitem[Amershi et~al\mbox{.}(2019)]%
        {Amershi2019-oh}
\bibfield{author}{\bibinfo{person}{Saleema Amershi}, \bibinfo{person}{Dan
  Weld}, \bibinfo{person}{Mihaela Vorvoreanu}, \bibinfo{person}{Adam Fourney},
  \bibinfo{person}{Besmira Nushi}, \bibinfo{person}{Penny Collisson},
  \bibinfo{person}{Jina Suh}, \bibinfo{person}{Shamsi Iqbal},
  \bibinfo{person}{Paul~N Bennett}, \bibinfo{person}{Kori Inkpen},
  \bibinfo{person}{Jaime Teevan}, \bibinfo{person}{Ruth Kikin-Gil}, {and}
  \bibinfo{person}{Eric Horvitz}.} \bibinfo{year}{2019}\natexlab{}.
\newblock \showarticletitle{Guidelines for {Human-AI} Interaction}. In
  \bibinfo{booktitle}{\emph{Proceedings of the 2019 {CHI} Conference on Human
  Factors in Computing Systems}} (Glasgow, Scotland Uk)
  \emph{(\bibinfo{series}{CHI '19}, \bibinfo{number}{Paper 3})}.
  \bibinfo{publisher}{Association for Computing Machinery},
  \bibinfo{address}{New York, NY, USA}, \bibinfo{pages}{1--13}.
\newblock
\showISBNx{9781450359702}
\urldef\tempurl%
\url{https://doi.org/10.1145/3290605.3300233}
\showDOI{\tempurl}


\bibitem[Bardzell et~al\mbox{.}(2014)]%
        {Bardzell2014-cx}
\bibfield{author}{\bibinfo{person}{Jeffrey Bardzell}, \bibinfo{person}{Shaowen
  Bardzell}, {and} \bibinfo{person}{Austin Toombs}.}
  \bibinfo{year}{2014}\natexlab{}.
\newblock \showarticletitle{``Now That's Definitely a Proper Hack'': Self-made
  Tools in Hackerspaces}. In \bibinfo{booktitle}{\emph{Proceedings of the 32nd
  annual {ACM} conference on Human factors in computing systems - {CHI} '14}}
  (Toronto, Ontario, Canada) \emph{(\bibinfo{series}{CHI '14})}.
  \bibinfo{publisher}{ACM Press}, \bibinfo{address}{New York, New York, USA},
  \bibinfo{pages}{473--476}.
\newblock
\showISBNx{9781450324731}
\urldef\tempurl%
\url{https://doi.org/10.1145/2556288.2557221}
\showDOI{\tempurl}


\bibitem[Biskjaer and Halskov(2014)]%
        {Biskjaer2014-gy}
\bibfield{author}{\bibinfo{person}{Michael~Mose Biskjaer} {and}
  \bibinfo{person}{Kim Halskov}.} \bibinfo{year}{2014}\natexlab{}.
\newblock \showarticletitle{Decisive constraints as a creative resource in
  interaction design}.
\newblock \bibinfo{journal}{\emph{Digital Creativity}} \bibinfo{volume}{25},
  \bibinfo{number}{1} (\bibinfo{date}{Jan.} \bibinfo{year}{2014}),
  \bibinfo{pages}{27--61}.
\newblock
\showISSN{1462-6268}
\urldef\tempurl%
\url{https://doi.org/10.1080/14626268.2013.855239}
\showDOI{\tempurl}


\bibitem[Bongard-Blanchy et~al\mbox{.}(2021)]%
        {Bongard-Blanchy2021-wj}
\bibfield{author}{\bibinfo{person}{Kerstin Bongard-Blanchy},
  \bibinfo{person}{Arianna Rossi}, \bibinfo{person}{Salvador Rivas},
  \bibinfo{person}{Sophie Doublet}, \bibinfo{person}{Vincent Koenig}, {and}
  \bibinfo{person}{Gabriele Lenzini}.} \bibinfo{year}{2021}\natexlab{}.
\newblock \showarticletitle{''I am Definitely Manipulated, Even When {I} am
  Aware of it. It's Ridiculous!'' - Dark Patterns from the {End-User}
  Perspective}. In \bibinfo{booktitle}{\emph{Designing Interactive Systems
  Conference 2021}} (Virtual Event USA) \emph{(\bibinfo{series}{DIS '21},
  Vol.~\bibinfo{volume}{1})}. \bibinfo{publisher}{ACM}, \bibinfo{address}{New
  York, NY, USA}, \bibinfo{pages}{763--776}.
\newblock
\showISBNx{9781450384766}
\urldef\tempurl%
\url{https://doi.org/10.1145/3461778.3462086}
\showDOI{\tempurl}


\bibitem[Boyd and Shilton(2021)]%
        {Boyd2021-sv}
\bibfield{author}{\bibinfo{person}{Karen~L Boyd} {and} \bibinfo{person}{Katie
  Shilton}.} \bibinfo{year}{2021}\natexlab{}.
\newblock \showarticletitle{Adapting Ethical Sensitivity as a Construct to
  Study Technology Design Teams}.
\newblock \bibinfo{journal}{\emph{Proc. ACM Hum.-Comput. Interact.}}
  \bibinfo{volume}{5}, \bibinfo{number}{GROUP} (\bibinfo{date}{July}
  \bibinfo{year}{2021}), \bibinfo{pages}{1--29}.
\newblock
\urldef\tempurl%
\url{https://doi.org/10.1145/3463929}
\showDOI{\tempurl}


\bibitem[Braun and Clarke(2021)]%
        {Braun2021-dt}
\bibfield{author}{\bibinfo{person}{Virginia Braun} {and}
  \bibinfo{person}{Victoria Clarke}.} \bibinfo{year}{2021}\natexlab{}.
\newblock \bibinfo{booktitle}{\emph{Thematic Analysis: A Practical Guide}}.
\newblock \bibinfo{publisher}{SAGE}.
\newblock
\showISBNx{9781526417299}
\urldef\tempurl%
\url{https://play.google.com/store/books/details?id=eMArEAAAQBAJ}
\showURL{%
\tempurl}


\bibitem[Burnett et~al\mbox{.}(2016)]%
        {Burnett2016-dr}
\bibfield{author}{\bibinfo{person}{Margaret Burnett}, \bibinfo{person}{Simone
  Stumpf}, \bibinfo{person}{Jamie Macbeth}, \bibinfo{person}{Stephann Makri},
  \bibinfo{person}{Laura Beckwith}, \bibinfo{person}{Irwin Kwan},
  \bibinfo{person}{Anicia Peters}, {and} \bibinfo{person}{William Jernigan}.}
  \bibinfo{year}{2016}\natexlab{}.
\newblock \showarticletitle{{GenderMag}: A Method for Evaluating Software's
  Gender Inclusiveness}.
\newblock \bibinfo{journal}{\emph{Interacting with computers}}
  \bibinfo{volume}{28}, \bibinfo{number}{6} (\bibinfo{date}{Oct.}
  \bibinfo{year}{2016}), \bibinfo{pages}{760--787}.
\newblock
\showISSN{0953-5438}
\urldef\tempurl%
\url{https://doi.org/10.1093/iwc/iwv046}
\showDOI{\tempurl}


\bibitem[Buwert(2018)]%
        {Buwert2018-uw}
\bibfield{author}{\bibinfo{person}{Peter Buwert}.}
  \bibinfo{year}{2018}\natexlab{}.
\newblock \showarticletitle{Examining the Professional Codes of Design
  Organisations}. In \bibinfo{booktitle}{\emph{Proceedings of the Design
  Research Society}}.
\newblock
\urldef\tempurl%
\url{https://doi.org/10.21606/dma.2017.493}
\showDOI{\tempurl}


\bibitem[Carspecken(1996)]%
        {Carspecken1996-bq}
\bibfield{author}{\bibinfo{person}{Phil~F Carspecken}.}
  \bibinfo{year}{1996}\natexlab{}.
\newblock \bibinfo{booktitle}{\emph{Critical ethnography in educational
  research: A theoretical and practical guide}}.
\newblock \bibinfo{publisher}{Routledge}, \bibinfo{address}{New York}.
\newblock
\showISBNx{9780415904933}


\bibitem[Chivukula et~al\mbox{.}(2021a)]%
        {Chivukula2021-oj}
\bibfield{author}{\bibinfo{person}{Shruthi~Sai Chivukula},
  \bibinfo{person}{Aiza Hasib}, \bibinfo{person}{Ziqing Li},
  \bibinfo{person}{Jingle Chen}, {and} \bibinfo{person}{Colin~M Gray}.}
  \bibinfo{year}{2021}\natexlab{a}.
\newblock \showarticletitle{Identity Claims that Underlie Ethical Awareness and
  Action}. In \bibinfo{booktitle}{\emph{Proceedings of the 2021 {CHI}
  Conference on Human Factors in Computing Systems}}
  \emph{(\bibinfo{series}{CHI'21})}.
\newblock
\urldef\tempurl%
\url{https://doi.org/10.1145/3411764.3445375}
\showDOI{\tempurl}


\bibitem[Chivukula et~al\mbox{.}(2021b)]%
        {chivukula2021identity}
\bibfield{author}{\bibinfo{person}{Shruthi~Sai Chivukula},
  \bibinfo{person}{Aiza Hasib}, \bibinfo{person}{Ziqing Li},
  \bibinfo{person}{Jingle Chen}, {and} \bibinfo{person}{Colin~M Gray}.}
  \bibinfo{year}{2021}\natexlab{b}.
\newblock \showarticletitle{Identity Claims that Underlie Ethical Awareness and
  Action}. In \bibinfo{booktitle}{\emph{Proceedings of the 2021 {CHI}
  Conference on Human Factors in Computing Systems}}
  \emph{(\bibinfo{series}{CHI'21})}.
\newblock
\urldef\tempurl%
\url{https://doi.org/10.1145/3411764.3445375}
\showDOI{\tempurl}


\bibitem[Chivukula et~al\mbox{.}(2021c)]%
        {Chivukula2021-xk}
\bibfield{author}{\bibinfo{person}{Shruthi~Sai Chivukula},
  \bibinfo{person}{Ziqing Li}, \bibinfo{person}{Anne~C Pivonka},
  \bibinfo{person}{Jingning Chen}, {and} \bibinfo{person}{Colin~M Gray}.}
  \bibinfo{year}{2021}\natexlab{c}.
\newblock \showarticletitle{Surveying the Landscape of {Ethics-Focused} Design
  Methods}.
\newblock  (\bibinfo{date}{Feb.} \bibinfo{year}{2021}).
\newblock
\showeprint[arxiv]{2102.08909}~[cs.HC]
\urldef\tempurl%
\url{http://arxiv.org/abs/2102.08909}
\showURL{%
\tempurl}


\bibitem[Chivukula et~al\mbox{.}(2020)]%
        {Chivukula2020-bv}
\bibfield{author}{\bibinfo{person}{Shruthi~Sai Chivukula},
  \bibinfo{person}{Chris~Rhys Watkins}, \bibinfo{person}{Rhea Manocha},
  \bibinfo{person}{Jingle Chen}, {and} \bibinfo{person}{Colin~M Gray}.}
  \bibinfo{year}{2020}\natexlab{}.
\newblock \showarticletitle{Dimensions of {UX} Practice that Shape Ethical
  Awareness}. In \bibinfo{booktitle}{\emph{Proceedings of the 2020 {CHI}
  Conference on Human Factors in Computing Systems}} (Honolulu, HI, USA)
  \emph{(\bibinfo{series}{CHI '20})}. \bibinfo{publisher}{Association for
  Computing Machinery}, \bibinfo{address}{New York, NY, USA},
  \bibinfo{pages}{1--13}.
\newblock
\showISBNx{9781450367080}
\urldef\tempurl%
\url{https://doi.org/10.1145/3313831.3376459}
\showDOI{\tempurl}


\bibitem[d'Aquin et~al\mbox{.}(2018)]%
        {DAquin2018-bl}
\bibfield{author}{\bibinfo{person}{Mathieu d'Aquin}, \bibinfo{person}{Pinelopi
  Troullinou}, \bibinfo{person}{Noel~E O'Connor}, \bibinfo{person}{Aindrias
  Cullen}, \bibinfo{person}{Gr{\'a}inne Faller}, {and} \bibinfo{person}{Louise
  Holden}.} \bibinfo{year}{2018}\natexlab{}.
\newblock \showarticletitle{Towards an Ethics by Design Methodology for {AI}
  Research Projects}. In \bibinfo{booktitle}{\emph{Proceedings of the 2018
  {AAAI/ACM} Conference on {AI}, Ethics, and Society}}.
  \bibinfo{pages}{54--59}.
\newblock
\urldef\tempurl%
\url{https://dl.acm.org/citation.cfm?id=3278765}
\showURL{%
\tempurl}


\bibitem[Debs et~al\mbox{.}(2022)]%
        {Debs2022-mt}
\bibfield{author}{\bibinfo{person}{Luciana Debs}, \bibinfo{person}{Colin~M
  Gray}, {and} \bibinfo{person}{Paul~A Asunda}.}
  \bibinfo{year}{2022}\natexlab{}.
\newblock \showarticletitle{Students' perceptions and reasoning patterns about
  the ethics of emerging technology}.
\newblock \bibinfo{journal}{\emph{International Journal of Technology and
  Design Education}} (\bibinfo{date}{Jan.} \bibinfo{year}{2022}).
\newblock
\showISSN{1573-1804}
\urldef\tempurl%
\url{https://doi.org/10.1007/s10798-021-09719-w}
\showDOI{\tempurl}


\bibitem[Dindler et~al\mbox{.}(2022)]%
        {Dindler2022-ny}
\bibfield{author}{\bibinfo{person}{Christian Dindler},
  \bibinfo{person}{Peter~Gall Krogh}, \bibinfo{person}{Kasper Tik{\ae}r}, {and}
  \bibinfo{person}{Peter N{\o}rreg{\aa}rd}.} \bibinfo{year}{2022}\natexlab{}.
\newblock \showarticletitle{Engagements and articulations of ethics in design
  practice}.
\newblock  (\bibinfo{year}{2022}).
\newblock
\urldef\tempurl%
\url{https://doi.org/10.57698/V16I2.04}
\showDOI{\tempurl}


\bibitem[Dorst(2015)]%
        {Dorst2015-aq}
\bibfield{author}{\bibinfo{person}{Kees Dorst}.}
  \bibinfo{year}{2015}\natexlab{}.
\newblock \showarticletitle{{Frame Creation and Design in the Expanded Field}}.
\newblock \bibinfo{journal}{\emph{She Ji: The Journal of Design, Economics, and
  Innovation}} \bibinfo{volume}{1}, \bibinfo{number}{1} (\bibinfo{date}{Jan.}
  \bibinfo{year}{2015}), \bibinfo{pages}{22--33}.
\newblock
\urldef\tempurl%
\url{http://linkinghub.elsevier.com/retrieve/pii/S2405872615300241}
\showURL{%
\tempurl}


\bibitem[Elsden et~al\mbox{.}(2017)]%
        {Elsden2017-lu}
\bibfield{author}{\bibinfo{person}{Chris Elsden}, \bibinfo{person}{David
  Chatting}, \bibinfo{person}{Abigail~C Durrant}, \bibinfo{person}{Andrew
  Garbett}, \bibinfo{person}{Bettina Nissen}, \bibinfo{person}{John Vines},
  {and} \bibinfo{person}{David~S Kirk}.} \bibinfo{year}{2017}\natexlab{}.
\newblock \showarticletitle{On Speculative Enactments}. In
  \bibinfo{booktitle}{\emph{Proceedings of the 2017 {CHI} Conference on Human
  Factors in Computing Systems}} (Denver, Colorado, USA)
  \emph{(\bibinfo{series}{CHI '17})}. \bibinfo{publisher}{Association for
  Computing Machinery}, \bibinfo{address}{New York, NY, USA},
  \bibinfo{pages}{5386--5399}.
\newblock
\showISBNx{9781450346559}
\urldef\tempurl%
\url{https://doi.org/10.1145/3025453.3025503}
\showDOI{\tempurl}


\bibitem[Flanagan and Nissenbaum(2014)]%
        {Flanagan2014-hf}
\bibfield{author}{\bibinfo{person}{Mary Flanagan} {and} \bibinfo{person}{Helen
  Nissenbaum}.} \bibinfo{year}{2014}\natexlab{}.
\newblock \bibinfo{booktitle}{\emph{Values at Play in Digital Games}}.
\newblock \bibinfo{publisher}{MIT Press}, \bibinfo{address}{Cambridge, MA}.
\newblock
\showISBNx{9780262027663}
\urldef\tempurl%
\url{https://market.android.com/details?id=book-iIYRBAAAQBAJ}
\showURL{%
\tempurl}


\bibitem[Fore and Hess(2019)]%
        {Fore2019-vi}
\bibfield{author}{\bibinfo{person}{Grant~A Fore} {and}
  \bibinfo{person}{Justin~L Hess}.} \bibinfo{year}{2019}\natexlab{}.
\newblock \showarticletitle{Operationalizing Ethical Becoming as a Theoretical
  Framework for Teaching Engineering Design Ethics}.
\newblock \bibinfo{journal}{\emph{Science and engineering ethics}}
  (\bibinfo{date}{Nov.} \bibinfo{year}{2019}).
\newblock
\showISSN{1353-3452, 1471-5546}
\urldef\tempurl%
\url{https://doi.org/10.1007/s11948-019-00160-w}
\showDOI{\tempurl}


\bibitem[Frauenberger et~al\mbox{.}(2017)]%
        {Frauenberger2017-mk}
\bibfield{author}{\bibinfo{person}{Christopher Frauenberger},
  \bibinfo{person}{Marjo Rauhala}, {and} \bibinfo{person}{Geraldine
  Fitzpatrick}.} \bibinfo{year}{2017}\natexlab{}.
\newblock \showarticletitle{{{In-Action} Ethics}}.
\newblock \bibinfo{journal}{\emph{Interacting with computers}}
  \bibinfo{volume}{29}, \bibinfo{number}{2} (\bibinfo{date}{March}
  \bibinfo{year}{2017}), \bibinfo{pages}{220--236}.
\newblock
\showISSN{0953-5438}
\urldef\tempurl%
\url{https://doi.org/10.1093/iwc/iww024}
\showDOI{\tempurl}


\bibitem[Friedman and Hendry(2019)]%
        {Friedman2019-zg}
\bibfield{author}{\bibinfo{person}{Batya Friedman} {and}
  \bibinfo{person}{David~G Hendry}.} \bibinfo{year}{2019}\natexlab{}.
\newblock \bibinfo{booktitle}{\emph{Value Sensitive Design: Shaping Technology
  with Moral Imagination}}.
\newblock \bibinfo{publisher}{MIT Press}.
\newblock
\showISBNx{9780262039536}
\urldef\tempurl%
\url{https://market.android.com/details?id=book-C4FruwEACAAJ}
\showURL{%
\tempurl}


\bibitem[Friedman et~al\mbox{.}(2017)]%
        {Friedman2017-rd}
\bibfield{author}{\bibinfo{person}{Batya Friedman}, \bibinfo{person}{David~G
  Hendry}, {and} \bibinfo{person}{Alan Borning}.}
  \bibinfo{year}{2017}\natexlab{}.
\newblock \bibinfo{booktitle}{\emph{A Survey of Value Sensitive Design
  Methods}}.
\newblock \bibinfo{publisher}{Now Publishers}.
\newblock
\showISBNx{9781680832907}
\urldef\tempurl%
\url{https://market.android.com/details?id=book-pW8etAEACAAJ}
\showURL{%
\tempurl}


\bibitem[Frost(2016)]%
        {Frost2016-jc}
\bibfield{author}{\bibinfo{person}{Brad Frost}.}
  \bibinfo{year}{2016}\natexlab{}.
\newblock \bibinfo{booktitle}{\emph{Atomic design}}.
\newblock \bibinfo{publisher}{Brad Frost Pittsburgh}.
\newblock
\urldef\tempurl%
\url{http://www.softouch.on.ca/kb/data/Atomic%20Design.pdf}
\showURL{%
\tempurl}


\bibitem[G.~Pillai et~al\mbox{.}(2021)]%
        {G_Pillai2021-hm}
\bibfield{author}{\bibinfo{person}{Ajit G.~Pillai}, \bibinfo{person}{A
  Baki~Kocaballi}, \bibinfo{person}{Tuck Wah~Leong}, \bibinfo{person}{Rafael
  A.~Calvo}, \bibinfo{person}{Nassim Parvin}, \bibinfo{person}{Katie Shilton},
  \bibinfo{person}{Jenny Waycott}, \bibinfo{person}{Casey Fiesler},
  \bibinfo{person}{John C.~Havens}, {and} \bibinfo{person}{Naseem Ahmadpour}.}
  \bibinfo{year}{2021}\natexlab{}.
\newblock \showarticletitle{Co-designing Resources for Ethics Education in
  {HCI}}.
\newblock In \bibinfo{booktitle}{\emph{Extended Abstracts of the 2021 {CHI}
  Conference on Human Factors in Computing Systems}}.
  \bibinfo{publisher}{Association for Computing Machinery},
  \bibinfo{address}{New York, NY, USA}, \bibinfo{pages}{1--5}.
\newblock
\showISBNx{9781450380959}
\urldef\tempurl%
\url{https://doi.org/10.1145/3411763.3441349}
\showDOI{\tempurl}


\bibitem[Gispen(2017)]%
        {ethicalcontract}
\bibfield{author}{\bibinfo{person}{Jet Gispen}.}
  \bibinfo{year}{2017}\natexlab{}.
\newblock \bibinfo{title}{Ethical Contract}.
\newblock
  \bibinfo{howpublished}{\url{https://www.ethicsfordesigners.com/ethical-contract}}.
\newblock


\bibitem[Goodman et~al\mbox{.}(2011)]%
        {Goodman2011-ak}
\bibfield{author}{\bibinfo{person}{Elizabeth Goodman}, \bibinfo{person}{Erik
  Stolterman}, {and} \bibinfo{person}{Ron Wakkary}.}
  \bibinfo{year}{2011}\natexlab{}.
\newblock \showarticletitle{Understanding Interaction Design Practices}. In
  \bibinfo{booktitle}{\emph{Proceedings of the {SIGCHI} Conference on Human
  Factors in Computing Systems}} (Vancouver, BC, Canada)
  \emph{(\bibinfo{series}{CHI '11})}. \bibinfo{publisher}{ACM},
  \bibinfo{address}{New York, NY, USA}, \bibinfo{pages}{1061--1070}.
\newblock
\showISBNx{9781450302289}
\urldef\tempurl%
\url{https://doi.org/10.1145/1978942.1979100}
\showDOI{\tempurl}


\bibitem[Gray(2016)]%
        {Gray2016-pa}
\bibfield{author}{\bibinfo{person}{Colin~M Gray}.}
  \bibinfo{year}{2016}\natexlab{}.
\newblock \showarticletitle{It's More of a Mindset Than a Method: {UX}
  Practitioners' Conception of Design Methods}. In
  \bibinfo{booktitle}{\emph{Proceedings of the 2016 {CHI} Conference on Human
  Factors in Computing Systems}} (Santa Clara, California, USA).
  \bibinfo{publisher}{ACM}, \bibinfo{address}{New York, New York, USA},
  \bibinfo{pages}{4044--4055}.
\newblock
\showISBNx{9781450333627}
\urldef\tempurl%
\url{https://doi.org/10.1145/2858036.2858410}
\showDOI{\tempurl}


\bibitem[Gray(2022)]%
        {Gray2022-na}
\bibfield{author}{\bibinfo{person}{Colin~M Gray}.}
  \bibinfo{year}{2022}\natexlab{}.
\newblock \showarticletitle{Languaging design methods}.
\newblock \bibinfo{journal}{\emph{Design Studies}}  \bibinfo{volume}{78}
  (\bibinfo{date}{Jan.} \bibinfo{year}{2022}), \bibinfo{pages}{101076}.
\newblock
\showISSN{0142-694X}
\urldef\tempurl%
\url{https://doi.org/10.1016/j.destud.2021.101076}
\showDOI{\tempurl}


\bibitem[Gray and Boling(2017)]%
        {Gray2017-dx}
\bibfield{author}{\bibinfo{person}{Colin~M Gray} {and}
  \bibinfo{person}{Elizabeth Boling}.} \bibinfo{year}{2017}\natexlab{}.
\newblock \showarticletitle{Designers' Articulation and Activation of
  Instrumental Design Judgments in {Cross-Cultural} User Research}.
\newblock In \bibinfo{booktitle}{\emph{Analysing Design Thinking: Studies of
  {Cross-Cultural} {Co-Creation}}}, \bibfield{editor}{\bibinfo{person}{Bo~T
  Christensen}, \bibinfo{person}{Linden~J Ball}, {and} \bibinfo{person}{Kim
  Halskov}} (Eds.). \bibinfo{publisher}{CRC Press}, \bibinfo{address}{Boca
  Raton, FL}, \bibinfo{pages}{191--214}.
\newblock


\bibitem[Gray and Chivukula(2019)]%
        {Gray2019-ep}
\bibfield{author}{\bibinfo{person}{Colin~M Gray} {and}
  \bibinfo{person}{Shruthi~Sai Chivukula}.} \bibinfo{year}{2019}\natexlab{}.
\newblock \showarticletitle{Ethical Mediation in {UX} Practice}. In
  \bibinfo{booktitle}{\emph{Proceedings of the 2019 {CHI} Conference on Human
  Factors in Computing Systems}} (Glasgow, Scotland Uk)
  \emph{(\bibinfo{series}{CHI '19}, \bibinfo{number}{Paper 178})}.
  \bibinfo{publisher}{Association for Computing Machinery},
  \bibinfo{address}{New York, NY, USA}, \bibinfo{pages}{1--11}.
\newblock
\showISBNx{9781450359702}
\urldef\tempurl%
\url{https://doi.org/10.1145/3290605.3300408}
\showDOI{\tempurl}


\bibitem[Gray et~al\mbox{.}(2021a)]%
        {Gray2021-gl}
\bibfield{author}{\bibinfo{person}{Colin~M Gray}, \bibinfo{person}{Shruthi~Sai
  Chivukula}, \bibinfo{person}{Kassandra Melkey}, {and} \bibinfo{person}{Rhea
  Manocha}.} \bibinfo{year}{2021}\natexlab{a}.
\newblock \showarticletitle{Understanding ``dark'' design roles in computing
  education}. In \bibinfo{booktitle}{\emph{Proceedings of the 17th {ACM}
  Conference on International Computing Education Research}} (Virtual Event
  USA). \bibinfo{publisher}{ACM}, \bibinfo{address}{New York, NY, USA}.
\newblock
\showISBNx{9781450383264}
\urldef\tempurl%
\url{https://doi.org/10.1145/3446871.3469754}
\showDOI{\tempurl}


\bibitem[Gray et~al\mbox{.}(2015)]%
        {Gray2015-qi}
\bibfield{author}{\bibinfo{person}{Colin~M Gray}, \bibinfo{person}{Cesur
  Dagli}, \bibinfo{person}{Muruvvet Demiral-Uzan}, \bibinfo{person}{Funda
  Ergulec}, \bibinfo{person}{Verily Tan}, \bibinfo{person}{Abdullah~A
  Altuwaijri}, \bibinfo{person}{Khendum Gyabak}, \bibinfo{person}{Megan
  Hilligoss}, \bibinfo{person}{Remzi Kizilboga}, \bibinfo{person}{Kei Tomita},
  {and} \bibinfo{person}{Elizabeth Boling}.} \bibinfo{year}{2015}\natexlab{}.
\newblock \showarticletitle{{Judgment and Instructional Design: How {ID}
  Practitioners Work In Practice}}.
\newblock \bibinfo{journal}{\emph{Performance Improvement Quarterly}}
  \bibinfo{volume}{28}, \bibinfo{number}{3} (\bibinfo{date}{Oct.}
  \bibinfo{year}{2015}), \bibinfo{pages}{25--49}.
\newblock
\showISSN{0898-5952}
\urldef\tempurl%
\url{https://doi.org/10.1002/piq.21198}
\showDOI{\tempurl}


\bibitem[Gray et~al\mbox{.}(2022)]%
        {Gray2022-kv}
\bibfield{author}{\bibinfo{person}{Colin~M Gray}, \bibinfo{person}{Aiza Hasib},
  \bibinfo{person}{Ziqing Li}, {and} \bibinfo{person}{Shruthi~Sai Chivukula}.}
  \bibinfo{year}{2022}\natexlab{}.
\newblock \showarticletitle{Using decisive constraints to create design methods
  that guide ethical impact}.
\newblock \bibinfo{journal}{\emph{Design Studies}}  \bibinfo{volume}{79}
  (\bibinfo{date}{March} \bibinfo{year}{2022}), \bibinfo{pages}{101097}.
\newblock
\showISSN{0142-694X}
\urldef\tempurl%
\url{https://doi.org/10.1016/j.destud.2022.101097}
\showDOI{\tempurl}


\bibitem[Gray et~al\mbox{.}(2021b)]%
        {Gray2021-zf}
\bibfield{author}{\bibinfo{person}{Colin~M Gray}, \bibinfo{person}{Cristiana
  Santos}, \bibinfo{person}{Nataliia Bielova}, \bibinfo{person}{Michael Toth},
  {and} \bibinfo{person}{Damian Clifford}.} \bibinfo{year}{2021}\natexlab{b}.
\newblock \showarticletitle{Dark Patterns and the Legal Requirements of Consent
  Banners: An Interaction Criticism Perspective}. In
  \bibinfo{booktitle}{\emph{Proceedings of the 2021 {CHI} Conference on Human
  Factors in Computing Systems}} \emph{(\bibinfo{series}{CHI'21})}.
  \bibinfo{publisher}{ACM Press}.
\newblock
\urldef\tempurl%
\url{https://doi.org/10.1145/3411764.3445779}
\showDOI{\tempurl}


\bibitem[Gross et~al\mbox{.}(2013)]%
        {Gross2013-qf}
\bibfield{author}{\bibinfo{person}{Shad Gross}, \bibinfo{person}{Tyler Pace},
  \bibinfo{person}{Jeffrey Bardzell}, {and} \bibinfo{person}{Shaowen
  Bardzell}.} \bibinfo{year}{2013}\natexlab{}.
\newblock \showarticletitle{Machinima production tools: a vernacular history of
  a creative medium}. In \bibinfo{booktitle}{\emph{Proceedings of the {SIGCHI}
  Conference on Human Factors in Computing Systems - {CHI} '13}} (Paris,
  France). \bibinfo{publisher}{ACM Press}, \bibinfo{address}{New York, New
  York, USA}, \bibinfo{pages}{971}.
\newblock
\showISBNx{9781450318990}
\urldef\tempurl%
\url{https://doi.org/10.1145/2470654.2466124}
\showDOI{\tempurl}


\bibitem[Hsieh and Shannon(2005)]%
        {Hsieh2005-ld}
\bibfield{author}{\bibinfo{person}{Hsiu-Fang Hsieh} {and}
  \bibinfo{person}{Sarah~E Shannon}.} \bibinfo{year}{2005}\natexlab{}.
\newblock \showarticletitle{Three approaches to qualitative content analysis}.
\newblock \bibinfo{journal}{\emph{Qualitative health research}}
  \bibinfo{volume}{15}, \bibinfo{number}{9} (\bibinfo{date}{Nov.}
  \bibinfo{year}{2005}), \bibinfo{pages}{1277--1288}.
\newblock
\showISSN{1049-7323}
\urldef\tempurl%
\url{https://doi.org/10.1177/1049732305276687}
\showDOI{\tempurl}


\bibitem[Lindberg et~al\mbox{.}(2020)]%
        {Lindberg2020-wk}
\bibfield{author}{\bibinfo{person}{Sharon Lindberg}, \bibinfo{person}{Petter
  Karlstr{\"o}m}, {and} \bibinfo{person}{Sirkku M{\"a}nnikk{\"o}~Barbutiu}.}
  \bibinfo{year}{2020}\natexlab{}.
\newblock \showarticletitle{Cultivating Ethics -- A perspective from practice}.
  In \bibinfo{booktitle}{\emph{Proceedings of the 11th Nordic Conference on
  {Human-Computer} Interaction: Shaping Experiences, Shaping Society}}
  (Tallinn, Estonia) \emph{(\bibinfo{series}{NordiCHI '20},
  \bibinfo{number}{Article 22})}. \bibinfo{publisher}{Association for Computing
  Machinery}, \bibinfo{address}{New York, NY, USA}, \bibinfo{pages}{1--11}.
\newblock
\showISBNx{9781450375795}
\urldef\tempurl%
\url{https://doi.org/10.1145/3419249.3420064}
\showDOI{\tempurl}


\bibitem[Lindberg et~al\mbox{.}(2021)]%
        {Lindberg2021-hi}
\bibfield{author}{\bibinfo{person}{Sharon Lindberg}, \bibinfo{person}{Petter
  Karlstr{\"o}m}, {and} \bibinfo{person}{Sirkku M{\"a}nnikk{\"o}~Barbutiu}.}
  \bibinfo{year}{2021}\natexlab{}.
\newblock \showarticletitle{Design Ethics in Practice - Points of Departure}.
\newblock \bibinfo{journal}{\emph{Proc. ACM Hum.-Comput. Interact.}}
  \bibinfo{volume}{5}, \bibinfo{number}{CSCW1} (\bibinfo{date}{April}
  \bibinfo{year}{2021}), \bibinfo{pages}{1--19}.
\newblock
\urldef\tempurl%
\url{https://doi.org/10.1145/3449204}
\showDOI{\tempurl}


\bibitem[Loke and Matthews(2020)]%
        {Loke2020-oe}
\bibfield{author}{\bibinfo{person}{Lian Loke} {and} \bibinfo{person}{Ben
  Matthews}.} \bibinfo{year}{2020}\natexlab{}.
\newblock \showarticletitle{Scaffolding of Interaction Design Education Towards
  Ethical Design Thinking}.
\newblock In \bibinfo{booktitle}{\emph{Design Thinking in Higher Education:
  Interdisciplinary Encounters}}, \bibfield{editor}{\bibinfo{person}{Gavin
  Melles}} (Ed.). \bibinfo{publisher}{Springer Singapore},
  \bibinfo{address}{Singapore}, \bibinfo{pages}{165--181}.
\newblock
\showISBNx{9789811557804}
\urldef\tempurl%
\url{https://doi.org/10.1007/978-981-15-5780-4\_8}
\showDOI{\tempurl}


\bibitem[Murdoch-Kitt et~al\mbox{.}(2020)]%
        {Murdoch-Kitt2020-sw}
\bibfield{author}{\bibinfo{person}{Kelly Murdoch-Kitt},
  \bibinfo{person}{Colin~M Gray}, \bibinfo{person}{Paul Parsons},
  \bibinfo{person}{Austin~L Toombs}, \bibinfo{person}{Marti Louw}, {and}
  \bibinfo{person}{Elona Van~Gent}.} \bibinfo{year}{2020}\natexlab{}.
\newblock \showarticletitle{Developing Students' Instrumental Judgment Capacity
  for Design Research Methods}. In \bibinfo{booktitle}{\emph{Dialogue:
  Proceedings of the {AIGA} Design Educators Community Conferences}},
  Vol.~\bibinfo{volume}{Decipher,1}. \bibinfo{publisher}{AIGA Design Educators
  Community}, \bibinfo{pages}{108--115}.
\newblock
\urldef\tempurl%
\url{https://doi.org/10.3998/mpub.11688977}
\showDOI{\tempurl}


\bibitem[Nelson and Stolterman(2012)]%
        {Nelson2012-ov}
\bibfield{author}{\bibinfo{person}{Harold~G Nelson} {and} \bibinfo{person}{Erik
  Stolterman}.} \bibinfo{year}{2012}\natexlab{}.
\newblock \bibinfo{booktitle}{\emph{The design way : Intentional change in an
  unpredictable world} (\bibinfo{edition}{2nd} ed.)}.
\newblock \bibinfo{publisher}{MIT Press}, \bibinfo{address}{Cambridge, MA}.
\newblock
\showISBNx{9780877783053}


\bibitem[Pivonka et~al\mbox{.}(2022)]%
        {Pivonka2022-nm}
\bibfield{author}{\bibinfo{person}{Anne Pivonka}, \bibinfo{person}{Laura
  Makary}, {and} \bibinfo{person}{Colin~M Gray}.}
  \bibinfo{year}{2022}\natexlab{}.
\newblock \showarticletitle{Organizing Metaphors for Design Methods in
  Intermediate {HCI} Education}. In \bibinfo{booktitle}{\emph{{EduCHI'22}: 4th
  Annual Symposium on {HCI} Education}}.
\newblock


\bibitem[Reijers(2019)]%
        {Reijers2019-sd}
\bibfield{author}{\bibinfo{person}{Wessel Reijers}.}
  \bibinfo{year}{2019}\natexlab{}.
\newblock \showarticletitle{Moving from value sensitive design to virtuous
  practice design}.
\newblock \bibinfo{journal}{\emph{Journal of Information, Communication and
  Ethics in Society}} \bibinfo{volume}{17}, \bibinfo{number}{2}
  (\bibinfo{date}{Jan.} \bibinfo{year}{2019}), \bibinfo{pages}{196--209}.
\newblock
\showISSN{1477-996X}
\urldef\tempurl%
\url{https://doi.org/10.1108/JICES-10-2018-0080}
\showDOI{\tempurl}
\showeprint{https://doi.org/10.1108/JICES-10-2018-0080}


\bibitem[Sch{\"o}n(1984)]%
        {Schon1984-oe}
\bibfield{author}{\bibinfo{person}{Donald~A Sch{\"o}n}.}
  \bibinfo{year}{1984}\natexlab{}.
\newblock \showarticletitle{Problems, frames and perspectives on designing}.
\newblock \bibinfo{journal}{\emph{Design Studies}} \bibinfo{volume}{5},
  \bibinfo{number}{3} (\bibinfo{date}{July} \bibinfo{year}{1984}),
  \bibinfo{pages}{132--136}.
\newblock
\showISSN{0142-694X}
\urldef\tempurl%
\url{https://doi.org/10.1016/0142-694X(84)90002-4}
\showDOI{\tempurl}


\bibitem[Shilton(2013)]%
        {Shilton2013-dq}
\bibfield{author}{\bibinfo{person}{Katie Shilton}.}
  \bibinfo{year}{2013}\natexlab{}.
\newblock \showarticletitle{{Values Levers: Building Ethics into Design}}.
\newblock \bibinfo{journal}{\emph{Science, technology \& human values}}
  \bibinfo{volume}{38}, \bibinfo{number}{3} (\bibinfo{date}{May}
  \bibinfo{year}{2013}), \bibinfo{pages}{374--397}.
\newblock
\showISSN{0162-2439}
\urldef\tempurl%
\url{https://doi.org/10.1177/0162243912436985}
\showDOI{\tempurl}


\bibitem[Shilton(2018)]%
        {Shilton2018-ws}
\bibfield{author}{\bibinfo{person}{Katie Shilton}.}
  \bibinfo{year}{2018}\natexlab{}.
\newblock \showarticletitle{Values and Ethics in {Human-Computer} Interaction}.
\newblock \bibinfo{journal}{\emph{Foundations and Trends\textregistered{}
  Human--Computer Interaction}} \bibinfo{volume}{12}, \bibinfo{number}{2}
  (\bibinfo{year}{2018}), \bibinfo{pages}{107--171}.
\newblock
\showISSN{1551-3955}
\urldef\tempurl%
\url{https://doi.org/10.1561/1100000073}
\showDOI{\tempurl}


\bibitem[Shilton and Anderson(2017)]%
        {Shilton2017-zu}
\bibfield{author}{\bibinfo{person}{Katie Shilton} {and} \bibinfo{person}{Sara
  Anderson}.} \bibinfo{year}{2017}\natexlab{}.
\newblock \showarticletitle{Blended, Not Bossy: Ethics Roles, Responsibilities
  and Expertise in Design}.
\newblock \bibinfo{journal}{\emph{Interacting with computers}}
  \bibinfo{volume}{29}, \bibinfo{number}{1} (\bibinfo{date}{Jan.}
  \bibinfo{year}{2017}), \bibinfo{pages}{71--79}.
\newblock
\showISSN{0953-5438}
\urldef\tempurl%
\url{https://doi.org/10.1093/iwc/iww002}
\showDOI{\tempurl}


\bibitem[Shilton et~al\mbox{.}(2021)]%
        {Shilton2021-gy}
\bibfield{author}{\bibinfo{person}{Katie Shilton}, \bibinfo{person}{Megan
  Finn}, {and} \bibinfo{person}{Quinn DuPont}.}
  \bibinfo{year}{2021}\natexlab{}.
\newblock \showarticletitle{Shaping ethical computing cultures}.
\newblock \bibinfo{journal}{\emph{Commun. ACM}} \bibinfo{volume}{64},
  \bibinfo{number}{11} (\bibinfo{date}{Oct.} \bibinfo{year}{2021}),
  \bibinfo{pages}{26--29}.
\newblock
\showISSN{0001-0782}
\urldef\tempurl%
\url{https://doi.org/10.1145/3486639}
\showDOI{\tempurl}


\bibitem[Slager et~al\mbox{.}(2021)]%
        {Slager2021-vv}
\bibfield{author}{\bibinfo{person}{Kyle Slager}, \bibinfo{person}{Ruby Nunez},
  \bibinfo{person}{William Short}, {and} \bibinfo{person}{Stacy~A Doore}.}
  \bibinfo{year}{2021}\natexlab{}.
\newblock \showarticletitle{Computing Ethics Starts on 'Day One': Ethics
  Narratives in Introductory {CS} Courses}. In
  \bibinfo{booktitle}{\emph{Proceedings of the 52nd {ACM} Technical Symposium
  on Computer Science Education}} (Virtual Event, USA)
  \emph{(\bibinfo{series}{SIGCSE '21})}. \bibinfo{publisher}{Association for
  Computing Machinery}, \bibinfo{address}{New York, NY, USA},
  \bibinfo{pages}{1282}.
\newblock
\showISBNx{9781450380621}
\urldef\tempurl%
\url{https://doi.org/10.1145/3408877.3439648}
\showDOI{\tempurl}


\bibitem[Stolterman(2008)]%
        {Stolterman2008-ho}
\bibfield{author}{\bibinfo{person}{E Stolterman}.}
  \bibinfo{year}{2008}\natexlab{}.
\newblock \showarticletitle{{The nature of design practice and implications for
  interaction design research}}.
\newblock \bibinfo{journal}{\emph{International Journal of Design}}
  \bibinfo{volume}{2}, \bibinfo{number}{1} (\bibinfo{date}{Jan.}
  \bibinfo{year}{2008}), \bibinfo{pages}{55--65}.
\newblock
\showISSN{0950-1991}
\urldef\tempurl%
\url{https://doi.org/10.1016/j.phymed.2007.09.005}
\showDOI{\tempurl}


\bibitem[Stolterman et~al\mbox{.}(2008)]%
        {Stolterman2008-ty}
\bibfield{author}{\bibinfo{person}{E Stolterman}, \bibinfo{person}{J McAtee},
  \bibinfo{person}{D Royer}, {and} \bibinfo{person}{S Thandapani}.}
  \bibinfo{year}{2008}\natexlab{}.
\newblock \showarticletitle{{Designerly tools}}. In
  \bibinfo{booktitle}{\emph{Undisciplined! Design Research Society Conference
  2008}} (Sheffield, UK). \bibinfo{publisher}{Sheffield Hallam University},
  \bibinfo{address}{Sheffield, UK}, \bibinfo{pages}{116:1--14}.
\newblock


\bibitem[Tulloch et~al\mbox{.}(2019)]%
        {Tulloch2019-dx}
\bibfield{author}{\bibinfo{person}{Angela Tulloch}, \bibinfo{person}{Tara
  French}, {and} \bibinfo{person}{Leigh-Anne Hepburn}.}
  \bibinfo{year}{2019}\natexlab{}.
\newblock \showarticletitle{Ethics by Design. Exploring Experiences of Harmony
  and Dissonance in Ethical Practice}.
\newblock \bibinfo{journal}{\emph{The Design Journal}} \bibinfo{volume}{22},
  \bibinfo{number}{sup1} (\bibinfo{date}{April} \bibinfo{year}{2019}),
  \bibinfo{pages}{401--416}.
\newblock
\showISSN{1460-6925}
\urldef\tempurl%
\url{https://doi.org/10.1080/14606925.2019.1595428}
\showDOI{\tempurl}


\bibitem[Watkins et~al\mbox{.}(2020)]%
        {Watkins2020-zr}
\bibfield{author}{\bibinfo{person}{Chris~Rhys Watkins},
  \bibinfo{person}{Colin~M Gray}, \bibinfo{person}{Austin~L Toombs}, {and}
  \bibinfo{person}{Paul Parsons}.} \bibinfo{year}{2020}\natexlab{}.
\newblock \showarticletitle{Tensions in Enacting a Design Philosophy in {UX}
  Practice}. In \bibinfo{booktitle}{\emph{{DIS'20}: Proceedings of the
  Designing Interactive Systems Conference 2020}}
  \emph{(\bibinfo{series}{DIS'20})}. \bibinfo{publisher}{ACM Press},
  \bibinfo{address}{New York, NY}.
\newblock
\urldef\tempurl%
\url{https://doi.org/10.1145/3357236.3395505}
\showDOI{\tempurl}


\bibitem[Wong(2021)]%
        {Wong2021-pv}
\bibfield{author}{\bibinfo{person}{Richmond~Y Wong}.}
  \bibinfo{year}{2021}\natexlab{}.
\newblock \showarticletitle{Tactics of Soft Resistance in User Experience
  Professionals' Values Work Tactics of Soft Resistance in User Experience
  Professionals' Values Work}.
\newblock \bibinfo{journal}{\emph{Proceedings of the ACM on Human-Computer
  Interaction}} \bibinfo{volume}{5}, \bibinfo{number}{CSCW2}
  (\bibinfo{date}{Oct.} \bibinfo{year}{2021}), \bibinfo{pages}{Article 355}.
\newblock
\urldef\tempurl%
\url{https://doi.org/10.1145/3479499}
\showDOI{\tempurl}


\bibitem[Wong et~al\mbox{.}(2020)]%
        {Wong2020-ki}
\bibfield{author}{\bibinfo{person}{Richmond~Y Wong}, \bibinfo{person}{Karen
  Boyd}, \bibinfo{person}{Jake Metcalf}, {and} \bibinfo{person}{Katie
  Shilton}.} \bibinfo{year}{2020}\natexlab{}.
\newblock \showarticletitle{Beyond Checklist Approaches to Ethics in Design}.
  In \bibinfo{booktitle}{\emph{Conference Companion Publication of the 2020 on
  Computer Supported Cooperative Work and Social Computing}} (Virtual Event,
  USA) \emph{(\bibinfo{series}{CSCW '20 Companion})}.
  \bibinfo{publisher}{Association for Computing Machinery},
  \bibinfo{address}{New York, NY, USA}, \bibinfo{pages}{511--517}.
\newblock
\showISBNx{9781450380591}
\urldef\tempurl%
\url{https://doi.org/10.1145/3406865.3418590}
\showDOI{\tempurl}


\bibitem[Woolrych et~al\mbox{.}(2011)]%
        {Woolrych2011-db}
\bibfield{author}{\bibinfo{person}{Alan Woolrych}, \bibinfo{person}{Kasper
  Hornb{\ae}k}, \bibinfo{person}{Erik Fr{\o}kj{\ae}r}, {and}
  \bibinfo{person}{Gilbert Cockton}.} \bibinfo{year}{2011}\natexlab{}.
\newblock \showarticletitle{{Ingredients and meals rather than recipes: A
  proposal for research that does not treat usability evaluation methods as
  indivisible wholes}}.
\newblock \bibinfo{journal}{\emph{International journal of human-computer
  interaction}} \bibinfo{volume}{27}, \bibinfo{number}{10}
  (\bibinfo{date}{Oct.} \bibinfo{year}{2011}), \bibinfo{pages}{940--970}.
\newblock
\showISSN{1044-7318, 1532-7590}
\urldef\tempurl%
\url{https://doi.org/10.1080/10447318.2011.555314}
\showDOI{\tempurl}


\bibitem[Yin(2009)]%
        {Yin2009-vs}
\bibfield{author}{\bibinfo{person}{Robert~K Yin}.}
  \bibinfo{year}{2009}\natexlab{}.
\newblock \bibinfo{booktitle}{\emph{Case study research : design and methods}}.
\newblock \bibinfo{publisher}{Sage Publications}, \bibinfo{address}{Los
  Angeles, Calif.} 312 pages.
\newblock
\showISBNx{9781412960991}
\urldef\tempurl%
\url{https://books.google.com/books?id=OgyqBAAAQBAJ&lr=&source=gbs_navlinks_s}
\showURL{%
\tempurl}


\end{thebibliography}

%%
%% If your work has an appendix, this is the place to put it.
%\appendix

\end{document}